# Enhanced Crystal Field Splitting and Orbital Selective Coherence by Strong Correlations in $V_2O_3$

A.I. Poteryaev,[1] J.M. Tomczak,[1] S. Biermann,[1] A. Georges,[1] A.I. Lichtenstein,[2] A.N. Rubtsov,[3] T. Saha-Dasgupta,[4] and O.K. Andersen[5]

[1] *Centre de Physique Théorique, Ecole Polytechnique, CNRS, 91128 Palaiseau, France*
[2] *I. Institut für Theoretische Physik, Universität Hamburg, Jungiusstraße 9, 20355 Hamburg, Germany*
[3] *Physics Department, Moscow State University, 119992 Moscow, Russia*
[4] *S. N. Bose. National Center for Basic Sciences JD Block, Salt Lake, Kolkata 700 098, India*
[5] *Max-Planck-Institut für Festkörperforschung, Heisenbergstraße 1, D-70569 Stuttgart, Germany*
(Dated: January 12, 2007)

We present a study of the paramagnetic metallic and insulating phases of vanadium sesquioxide by means of the $N$th order muffin-tin orbital implementation of density functional theory combined with dynamical mean-field theory. The transition is shown to be driven by a correlation-induced enhancement of the crystal field splitting within the $t_{2g}$ manifold, which results in a suppression of the hybridization between the $a_{1g}$ and $e_g^\pi$ bands. We discuss the changes in the effective quasi-particle band structure caused by the correlations and the corresponding self-energies. At temperatures of about 400 K we find the $a_{1g}$ orbitals to display coherent quasi-particle behavior, while a large imaginary part of the self-energy and broad features in the spectral function indicate that the $e_g^\pi$ orbitals are still far above their coherence temperature. The local spectral functions are in excellent agreement with recent bulk sensitive photoemission data. Finally, we also make a prediction for angle-resolved photoemission experiments by calculating momentum-resolved spectral functions.

## I. INTRODUCTION

The phase diagram of vanadium sesquioxide has attracted attention of solid-state physicists for decades[1]. Pure, corundum-structured $V_2O_3$ displays a paramagnetic metallic (PM) phase at ambient pressure and temperature. Cr-doping effectively expands the (ab) lattice constant and drives the paramagnetic metal into a paramagnetic insulating (PI) phase without changing the crystal symmetry. Application of physical pressure to the Cr-doped compound allows to recover the PM phase, thus leading to a unified phase diagram, in which Cr-doping can be considered equivalent to applying negative pressure. The PM and PI phases are separated by a first order line, which ends in a critical point at about 400 K. Upon cooling below 150 K, long-range antiferromagnetic order sets in, the structure distorts, and the compound becomes insulating (AFI phase). Doping with Ti impurities, on the other hand, corresponds to positive pressure. With increasing Ti concentration, the Néel temperature decreases and the AFI phase disappears at about 6% Ti-doping.

Doped $V_2O_3$ thus belongs to the class of materials in which a PM→PI transition can be studied by changing external parameters such as physical or chemical pressure. Numerous experimental studies have led to a detailed picture of the properties: The first order PM→PI transition is signaled by (i) a spectacular resistivity jump of seven orders of magnitude[2,3] (ii) a substantial transfer of spectral weight in the optical conductivity[4] from the low energy ($< 0.1$ eV) part of the spectrum to higher energies and (iii) the opening of a gap in the photoemission spectrum[5–9].

However, already the PM phase is characterized by strong Coulomb-correlation effects: bad metallic behavior of the resistivity, $\rho = 2 \times 10^{-4}$ $\Omega$ cm at ambient temperature[3], strong incoherent features in the optical[4] and photoemission spectra[8], and a strong mass enhancement as determined from specific heat measurements[10].

On the theoretical side, different complementary approaches – ranging from electronic structure calculations via theoretical models for the Mott transition to theories combining the two aspects – have been used to access the physics of the rich phase diagram:

McWhan *et al.*[2] emphasized the importance of on-site correlation effects, regarding $V_2O_3$ as a prototype for a Mott-Hubbard transition. Castellani *et al.*[11] studied a model based on the vanadium $t_{2g}$ Wannier functions split by the trigonal field into $a_{1g}$ and $e_g^\pi$ orbitals, and they discussed the possibility that in the ground state, one of the V electrons enters a bonding, molecular-orbital state between the $a_{1g}$ orbitals on a vertical V-pair, and the other electron enters the doubly degenerate $e_g^\pi$ level, thus giving rise to a spin $S = 1/2$ on each vanadium site. With the advent of dynamical mean-field theory (DMFT), Rozenberg *et al.*[12] investigated the similarities in the phenomenology of the one-band Hubbard model with the phase diagram of $V_2O_3$. The situation changed when polarization-dependent X-ray-absorption measurements of Park *et al.*[13] indicated that the spin is actually $S = 1$ in all $V_2O_3$ phases. This triggered refinements of the theoretical models[14,15].

Electronic-structure calculations by Mattheiss[16] using density-functional theory within the local density approximation (LDA) provided a parameter-free description of the chemical interactions, but they neither cap-tured the insulating nature of the PI phase nor the effects of strong correlations in the PM phase. Nevertheless, by addition of a few on-site Coulomb-interaction parameters, the LDA forms an accurate starting point for predictive calculations. In particular, the LDA+U (unrestricted Hartree-Fock, or static mean-field theory) calculations by Ezhov et al.[17] were the first to describe correctly the insulating nature of the low-temperature, antiferromagnetic, monoclinic phase. They found the ground state to have essentially the Hund's-rule configuration $e_g^{\pi\uparrow\uparrow}$, i.e., the $a_{1g}$ band to be empty. Subsequently, Elfimov et al.[18] demonstrated that the shape of the empty $a_{1g}^{\uparrow}$ band is determined not only by the hopping between the $a_{1g}$-orbitals of a vertical V-pair, but also by the many horizontal hopping integrals, and Eyert et al.[19] showed that the bonding part of the LDA $a_{1g}$ band is strongly hybridized with the $e_g^{\pi}$ band. All these observations led to a severe questioning of the $S = 1/2$ picture based on the bonding-antibonding splitting of the $a_{1g}$ band, in favor of the $S = 1$ picture with two $e_g^{\pi}$ electrons and an almost empty $a_{1g}$. This point of view was forcefully advocated by Sawatzky[23] and coworkers.

Combined with the LDA, realistic DMFT calculations for $V_2O_3$ were first performed by Held et al.[20,21]. As LDA input, they used the local density of states (DOS) projected onto respectively the $a_{1g}$ and $e_g^{\pi}$ partial waves, truncated outside the vanadium atomic sphere. In this kind of calculation[22], only the $a_{1g}$ and $e_g^{\pi}$ occupations are allowed to adjust through the DMFT iterations towards self-consistency. In an implementation of LDA+DMFT, employing an LDA Hamiltonian, also the orbital characters of each state may adjust, and this, we shall see, is crucial in vanadium sesquioxide where correlations lead to strong changes in the effective hybridizations between the $a_{1g}$ and $e_g^{\pi}$ orbital components, an effect which we shall call "correlation-induced re-hybridization" in the following. The importance of this effect was emphasized by Sawatzky[23], on the basis of the LDA+U calculations for the AFI phase[17]. Nevertheless, with LDA+DMFT[20,21] it was possible to obtain high-temperature spectral functions which exhibit lower and upper Hubbard peaks, a gap for the PI phase, and a quasi-particle peak for the PM phase in qualitative agreement with photoemission spectra (PES). However, improvement in the bulk sensitivity of PES for the PM phase at 175 K revealed a quasi-particle peak significantly wider and with more weight than the theoretical one[8]. Using a basis set of $t_{2g}$ Wannier functions, Anisimov et al.[24] performed Hamiltonian-based LDA+DMFT calculations and observed some changes from their earlier results[20,21]. However, they did not present a detailed comparison with the improved PES data of[8,9]. Laad et al.[26] emphasized the role of the crystal-field splitting and studied its influence on the PM-PI transition. Their LDA+DMFT implementation used a density of states approximation and iterative perturbation theory for solving the DMFT equations. In contrast to other LDA+DMFT work, they found the trigonal splitting to be screened out in the PM phase and, as a consequence, the quasi-particle peak to have exclusively $a_{1g}$ character.

In this paper we use the LDA $t_{2g}$ Wannier functions presented in detail in the companion paper (Ref.[27]), referred to as I, in order to construct a Hubbard Hamiltonian, which we solve in the dynamical mean-field approximation. A special effort is made in analytically continuing the quantum Monte Carlo data to the real-frequency axis. This allows in particular for a determination of the *correlated* (quasi-particle) band structure of the PM phase. We find excellent agreement with the improved PES data, which encourages us to present and discuss our momentum-resolved spectral function for comparison with future angle-resolved PES (ARPES). Our results reveal three key effects of strong correlations, which are responsible for the nature of the low energy behavior. These are i) a strongly orbital dependent coherence scale, ii) a considerable enhancement of the effective trigonal splitting between the $a_{1g}$ and the $e_g^{\pi}$ orbitals, to an extent such that the $a_{1g}$ band is nearly empty and iii) a strong reduction of the low energy bandwidths, associated with a small quasi-particle spectral weight for the $a_{1g}$ orbital. As a result, correlation effects considerably re-normalize downwards the effective low energy hybridization between the $a_{1g}$ and $e_g^{\pi}$ orbitals. In addition, higher energy structures in the self-energy lead to the expected lower and upper Hubbard bands. These findings provide quantitative support, already for the PM phase, for the $S = 1$ picture of two Hund's rule coupled electrons in the $e_g^{\pi}$ orbital, as in the AFI phase[17]. We also show that a slight lattice expansion decreases the $e_g^{\pi}$-$a_{1g}$ band overlap, and that because of the strong correlation-induced enhancement of the crystal field splitting, this leads to the opening of the gap associated with the PI phase.

In paper I, the details of the corundum structure, the LDA bands, the Wannier functions and their Hamiltonian generated by the $N$th-order muffin-tin orbital ($N$MTO) method[28] are given. Section II of the present paper briefly reviews the $N$MTO-implementation of LDA+DMFT. Section III presents our results. In section III A we discuss correlation effects on the low energy band structure of the paramagnetic phase, showing in particular how a correlation-enhanced crystal field splitting between the $a_{1g}$ and $e_g^{\pi}$ bands yields an effective suppression of the hybridization and a nearly empty $a_{1g}$ band. Section III B compares the resulting spectral function to recent photoemission experiments, with excellent agreement. The temperature-dependence of both, one-particle quantities and local susceptibilities is discussed in Section III C. Section III D is devoted to a discussion of the mechanism for the metal insulator transition, in particular regarding the role of the correlation-induced enhancement of the crystal field splitting. Results for the spectral function in the insulating phase are presented in Section III F. The Appendices give a detailed summary of both, the multi-orbital Quantum Monte Carlo algorithm by Hirsch and Fye and the continuous time

Quantum Monte Carlo algorithm by Rubtsov.

## II. METHOD

### A. Construction of the Wannier Hamiltonian

We construct a realistic many-electron Hamiltonian as the sum of a one-electron part and an on-site Coulomb repulsion and solve it in the dynamical mean-field approximation. The on-site repulsion depends on the representation (see below) and since we want the Hamiltonian to be both realistic and manageable, we use localized Wannier functions for a low energy subset of LDA bands, the vanadium $t_{2g}$ bands in the present case. This approach was first described in Ref.[29].

Our LDA Wannier functions are constructed as a symmetrically orthonormalized, minimal set of $N$th-order muffin-tin orbitals ($N$MTOs)[28]. The three $t_{2g}$ NMTOs for $V_2O_3$ are shown in Fig. 10 of paper I (Fig. I.10, in short). Such a basis function has *either* $a_{1g}$, or $e_{g,1}^\pi$, or $e_{g,2}^\pi$ character on the central vanadium site. The $N$MTOs on a site transform according to the irreducible representations of the point group, not of the full rotation group like atomic orbitals. For that reason, on-site matrices like the local self-energy and Green's function are diagonal in the $N$MTO representation. Moreover, an $N$MTO is maximally localized, in the sense that it has *no* $t_{2g}$ character on any *other* vanadium site. This localization criterion is natural for atom-centered orbitals, but is different from the one used for plane waves which minimizes the spread of the Wannier function[30]. For $3d$-systems, however, these two different localization criteria yield very similar Wannier functions[31].

In order for the basis set consisting of the three $t_{2g}$ $N$MTOs on each vanadium in the solid to span the $t_{2g}$-bands, each basis function must be a solution of Schrödinger's (or Dirac's) differential equation for the single-particle potential, $V^{LDA}$, for all (in practice $N+1$) energies in the band. This condition fixes all the other partial-wave characters of the $t_{2g}$ $N$MTOs, that is, *the oxygen and the non-$t_{2g}$ vanadium characters*. The result is seen in Fig. I.10, where in particular the oxygen $2p$ and the $3d\left(e_g^\sigma\right)$ characters on neighboring vanadiums are seen.

The $N$MTO overlap and Hamiltonan matrices, $\langle \chi_{Rm} \mid \chi_{R'm'} \rangle$ and $\langle \chi_{Rm} \left| -\Delta + V^{LDA} \right| \chi_{R'm'} \rangle$, with $R$ labelling the vanadium sites and $m$ the $t_{2g}$ orbitals, as well as the expansion coefficients of the $N$MTOs in partial waves, are all given in terms of the Korringa-Kohn-Rostoker Green matrix, downfolded to $t_{2g}$-space and evaluated at the chosen mesh of $N+1$ energies in the $t_{2g}$ band[28]. As may be seen from the bottom panel of Fig. I.5, the $t_{2g}$ bands were reproduced with merely $N=2$. Symmetrical orthonormalization finally yields the $N$MTO Wannier functions, $\chi_{Rm}^\perp(\mathbf{r} - \mathbf{R})$, and their Hamiltonian matrix,

$$\langle \chi_{Rm}^\perp \left| -\Delta + V^{LDA} \right| \chi_{R'm'}^\perp \rangle \equiv H_{Rm,R'm'}^{LDA}$$

whose elements are tabulated in Tables I.I and I.II. Note that $m = (a_{1g}, e_{g,1}^\pi, e_{g,2}^\pi)$ refers to irreducible partner functions of the point group of the vanadium site in question and, hence, to a *local* coordinate system which follows the symmetry operations between the four equivalent vanadium sites in the translational cell.

The complete Hamiltonian of the material is then constructed by supplementing the LDA $t_{2g}$ Hamiltonian with on-site Coulomb interactions:

$$\hat{H} = \hat{H}^{int} + \sum_{Rm,R'm',\sigma} H_{Rm,R'm'}^{LDA} \hat{c}_{Rm}^{\sigma\,\dagger} \hat{c}_{R'm'}^\sigma. \quad (1)$$

Here, $\hat{c}_{Rm}^{\sigma\,\dagger}$ [$\hat{c}_{Rm}^\sigma$] denotes the creation [annihilation] operator for an electron at site $R$ in orbital $m$ with spin $\sigma$, and

$$\hat{H}^{int} = \frac{1}{2} \sum_{R,m,m',\sigma} U_{mm'} \hat{n}_{Rm}^\sigma \hat{n}_{Rm'}^{-\sigma} + \quad (2)$$
$$\frac{1}{2} \sum_{R,m \neq m',\sigma} (U_{mm'} - J_{mm'}) \hat{n}_{Rm}^\sigma \hat{n}_{Rm'}^\sigma$$

is of Hubbard-Hund type, but can in principle contain more general interactions, such as spin-flip and pair-hopping or non-local terms. Here, $\hat{n}_{Rm}^\sigma = \hat{c}_{Rm}^{\sigma\,\dagger} \hat{c}_{Rm}^\sigma$ is the particle number operator and $U_{mm'}$ and $J_{mm'}$ are Coulomb and exchange integrals, which for $t_{2g}$ electrons can be parametrized[32] as follows: $U_{mm} = U$ is the Coulomb repulsion between electrons in the same orbital, $U_{m \neq m'} = U - 2J \equiv U'$ is the average repulsion, and $J_{m \neq m'} = J$ is the Hund's rule coupling. The latter depends only weakly on electronic screening in the solid and thus varies relatively little from compound to compound; for transition metal oxides values of about 0.7 eV have been estimated[13,34] and this is the value we used. In this work, we restrict ourselves to the form of the interaction given in Eq. (2), which allows us to use the QMC algorithm by Hirsch and Fye.

### B. Dynamical mean-field treatment

The resulting Hamiltonian (1) is then treated within dynamical mean-field theory, i.e. the self-energy is approximated by the self-energy of an effective local impurity problem embedded in an energy-dependent bath. This effective system is defined in terms of an effective action

$$S_{eff} = -\int_0^\beta d\tau \int_0^\beta d\tau' \sum_{m,m',\sigma} c_m^{\sigma\dagger}(\tau)[\mathcal{G}^{-1}(\tau-\tau')]_{mm'}^\sigma c_{m'}^\sigma(\tau') +$$

$$\frac{1}{2}\int_0^\beta d\tau \left\{ \sum_{m,m',\sigma} U_{mm'} n_m^\sigma(\tau) n_{m'}^{-\sigma}(\tau) + \sum_{m\neq m',\sigma} \left(U_{mm'} - J_{mm'}\right) n_m^\sigma(\tau) n_{m'}^\sigma(\tau) \right\}, \quad (3)$$

involving the *bath Green's function* $\mathcal{G}$ which describes the energy, orbital, spin and temperature dependent interactions of a given correlated atom with the rest of the crystal. $c_m^{\sigma\dagger}(\tau)$ and $c_m^\sigma(\tau)$ are Grassmann variables associated with the creation and annihilation operators introduced above. The bath Green's function has to be determined in a self-consistent way, following the general spirit of a mean-field theory.

The solution of the effective problem implies the calculation of the impurity Green's function

$$G_{mm'}^{loc}(\tau-\tau') = -\langle \mathcal{T} c_m^\sigma(\tau) c_{m'}^{\sigma+}(\tau') \rangle_{S_{eff}}, \quad (4)$$

where the subscript indicates that the average is to be performed using the effective action (3). This step is the most difficult within dynamical mean-field calculations. Different "impurity solvers" have been designed for this purpose. In the present paper we use the multi-band Quantum Monte Carlo schemes by Hirsch-Fye[49] and by Rubtsov[54], which – in contrast to approximate solver schemes such as iterative perturbation theory[26,35] or schemes based on the non-crossing approximation[36] – have the advantage of yielding numerically exact solutions to the problem. The algorithms are summarized in the Appendices A and B.

In the present case, $G^{loc}$ and the self-energy are diagonal in $m$ because on-site $N$MTO matrices transform according to the irreducible representations of the point group. Therefore all local quantities will carry only one orbital index in the following.

Once the impurity Green's function is known, the self-energy of the effective problem can be obtained as

$$\Sigma_m(\omega) = \mathcal{G}_m(\omega)^{-1} - G_m^{loc}(\omega)^{-1}, \quad (5)$$

for the imaginary Matsubara frequencies, i.e. for $\omega = i\omega_n = i(2n+1)\pi kT$. The dynamical mean-field approximation now amounts to using the self-energy $\Sigma_m$ of the effective problem as an approximation to the full non-local self-energy of the solid. That is,

$$-G_m^{loc}(\omega) \equiv -G_{mm}^{loc}(\omega) = \quad (6)$$
$$\left[H^{LDA} - \mu + \Sigma(\omega) - \omega\right]^{-1}_{Rm,Rm},$$

where the elements of the matrix to be inverted are

$$H^{LDA}_{Rm,R'm'} + [\Sigma_m(\omega) - \omega - \mu]\delta_{mm'}\delta_{RR'} \quad (7)$$
$$= -G^{-1}(\omega)_{Rm,R'm'}.$$

We obtain the 3×3 on-site Green matrix (6) by first Fourier-summing $\sum_T \exp(i\mathbf{k}\cdot\mathbf{T})$ the secular matrix (7) over the lattice translations, $\mathbf{T} = \mathbf{R} - \mathbf{R}'$, to the Bloch representation yielding,

$$H^{LDA}_{Im,I'm'}(\mathbf{k}) + [\Sigma_m(\omega) - \omega - \mu]\delta_{mm'}\delta_{II'} \quad (8)$$
$$= -G^{-1}(\mathbf{k},\omega)_{Im,I'm'},$$

in which $I$ and $I'$ label the 4 vanadium sites in the primitive cell. This 12×12 matrix (8) is then inverted for each $\mathbf{k}$ to give $G(\mathbf{k},i\omega_n)$. Finally, one of its 4 on-site blocks is integrated over $\mathbf{k}$ to form the local 3×3 Green matrix (6). Since the latter is diagonal, it suffices to calculate the diagonal $a_{1g}$, $e_{g,1}^\pi$ and $e_{g,2}^\pi$ elements.

In order to close the set of equations in a self-consistent way, the effective bath propagator is determined from Dyson's equation using the local Green's function of the solid which – according to DMFT – is required to equal the Green's function of the local effective problem. Iteration of this loop yields the dynamical mean-field solution of the Hamiltonian (1).

Finally, a few remarks:

(i) *Choice of local orbitals and of basis set.* As stressed above, the local Green's function $G^{loc}$, as well as the local self-energy $\Sigma$ and the local, energy-dependent effective Weiss field $\mathcal{G}^{-1}$ are expressed in the $t_{2g}$ Wannier representation. These quantities are thus (diagonal) 3×3 matrices in orbital space. The necessity of introducing a set of localized orbitals is intimately related to the dynamical mean-field concept: indeed, the notion of locality which has a natural meaning in the context of *lattice models* of correlated fermions does not have a basis-independent sense in the solid. However, as soon as a set of localized orbitals has been specified, the distinction between local and non-local quantities can in an obvious way be replaced by the distinction between on- and off-site ones. The latter notion acquires a well-defined meaning within classical muffin-tin or atomic-sphere based methods, but can in fact also be realized within different sets of Wannier functions or atomic-like orbitals. We stress that the orbitals in which the impurity model is expressed do not have to coincide with the basis set used for the implementation of the DMFT self-consistency condition. It is for technical convenience that here we use the $N$MTO basis set for both. For a general discussion of these points see[31].

(ii) *Double-counting correction.* In general, a double counting correction has to be subtracted from the LDA

Hamiltonian once Coulomb interactions are treated explicitly. In the present case of an effective Hamiltonian only containing correlated states, an important simplification occurs. As the double counting term is orbital-independent, it amounts to a mere shift of the origin of the energy and can thus be absorbed into the chemical potential $\mu = \mu_{true} + \zeta_{dc}$. In practice, it is then sufficient to determine the chemical potential in such a way that the correct particle number is obtained.

$$\begin{aligned} \hat{H} - \mu_{true}\hat{N} &= \hat{H}^{LDA} - \zeta_{dc}\hat{N} - \mu_{true}\hat{N} \\ &= \hat{H}^{LDA} - \mu\hat{N} \end{aligned}$$

(iii) *Self-consistent local density.* For the sake of simplicity we restricted the above description to the one-shot LDA+DMFT approach, in which the LDA Hamiltonian is calculated once and for all at the beginning. However, the method is in principle not restricted to this case: it is possible to recalculate at the end of the converged DMFT cycle the density of the system, to determine the $N$MTO-LDA Hamiltonian corresponding to this new density and to iterate the above procedure for this new Hamiltonian. By iterating until global self-consistency is reached one obtains an $N$MTO implementation of the full LDA+DMFT scheme. While in the $N$MTO context this has so far not been implemented, analogous calculations have been done in the LMTO and KKR framework[33].

(iv) *Choice of the U-matrix.* As mentioned above, the interaction vertex which supplements the LDA Hamiltonian can be parametrized in terms of two parameters, $U$ and $J$, corresponding to the intra-orbital Coulomb interaction and the Hund's-rule coupling, respectively. Whereas $J = 0.7$ eV is a generally accepted value for an early transition-metal oxide (hardly changed from its atomic value), the precise determination of $U$ is a delicate issue. The Coulomb interaction not only depends on the choice of representation, but also crucially on the electronic structure of the compound because $U$ is strongly screened. For metallic systems, ab initio techniques such as constrained LDA give only rough estimates. As a consequence, we treat $U$ as an adjustable parameter. With the LDA Hamiltonian calculated for the crystal structure of pure $V_2O_3$ at room temperature in I, we found (see next section) the LDA+DMFT solution to be insulating for $U > U_c \sim 4.35$ eV, i.e. for $U'_c \sim 2.95$ eV, and the best agreement with PM experimental data to be obtained with $U = 4.2$ eV, i.e. with $U' = 2.8$ eV. With the same value of $U$, we also correctly describe the insulating behavior of Cr-doped $V_2O_3$.

### C. Calculation of real frequency quantities

Angle-resolved photoemission spectroscopy (ARPES) measures $A(\mathbf{k}, \omega) = -\frac{1}{\pi}\mathrm{Tr}\Im G(\mathbf{k}, \omega + i0^+)$, so in order to predict such spectra we need to analytically continue the self-energies $\Sigma_m(i\omega_n)$ from imaginary to real frequencies. We do this by first analytically continuing the local Green's function obtained from a QMC calculation by using the maximum entropy algorithm[37]. The latter yields the local spectral function, $A_m(\omega) = \sum_{\mathbf{k}} A_m(\mathbf{k}, \omega)$ from which the local Green's function on the real-frequency axis is reconstructed by Kramers-Kronig transformation:

$$G_m(\omega + i0^+) = \int d\omega' \frac{A_m(\omega')}{\omega - \omega'} - i\pi A_m(\omega).$$

Finally, we use a root-finding algorithm to solve the coupled equations

$$\sum_{\mathbf{k}} \left[ \omega + \mu - \hat{H}^{LDA}(k) - \hat{\Sigma}(\omega) \right]^{-1}_{mm} = G_m(\omega)$$

for the complex quantities $\Sigma_m(\omega)$. Here, $G_m(\omega)$ is the matrix element of the on-site diagonal Green's function in Eq. (8). In order for this procedure to yield reliable solutions, well-converged Monte Carlo data with low statistical noise are needed. Insufficient statistics immediately manifests itself in yielding spiky self-energies. From our data we were able to obtain perfectly smooth self-energies (see e.g. Fig. 1), for which we checked that the corresponding $\mathbf{k}$-resolved spectral functions sum up to the total spectral function of Fig. 4. As a further test, we verified that we recover the original data on the imaginary-frequency axis from a Hilbert transform of the real-frequency self-energy data. Below we analyze both the orbital-resolved self-energies and the $\mathbf{k}$-resolved spectral functions obtained from them. For more details about the method, see[38]. For previous work on analytical continuations of self-energies see[24,40–42], and for early techniques to calculate $\mathbf{k}$-resolved spectra see[43,44].

## III. RESULTS

### A. Correlated paramagnetic metallic phase: self-energies, quasi-particle band structure and Hubbard bands

#### 1. A reminder on LDA bands

In order to understand how correlation effects deeply change the low energy band structure as compared to the LDA, we briefly recall here the essentials of $H^{LDA}(\mathbf{k})$, as obtained in paper I.

The LDA $t_{2g}$ bands, $E_j(\mathbf{k})$, are shown in the upper, left-most part of Fig. I.9. They are seen to extend from $\min E_{t_{2g}} \approx -1.0$ eV below to $\max E_{t_{2g}} \approx 1.5$ eV above the Fermi level and are thus substantially broader than the quasi-particle peak in Fig. 4, and substantially smaller than the distance $\sim U$ between the Hubbard bands. The LDA bands exhibit $a_{1g}$-$e_g^\pi$ hybridization, as can be seen by the projection onto the $a_{1g}$ and $e_g^\pi$ Wannier functions in Fig. I.8: Bands with both $a_{1g}$ and $e_g^\pi$ character are hybridized, and such bands are in particular found at the bottom of the $t_{2g}$ band near the L and F points[19]. A key quantity is the LDA crystal-field splitting, $\Delta_{\mathrm{LDA}}$=0.27 eV, listed in Table I.I, which is an order of magnitude smaller than the $t_{2g}$ bandwidth, and of the same size as each of the large $a_{1g}$-$e_g^\pi$ integrals for hopping to the three nearest neighbors in the $xy$-plane (Table I.II and bottom of Fig. I.12).

In the left-most bottom part of Fig. I.9, we show the bands, $\epsilon_{mj}(\mathbf{k})$, which result if all hopping integrals between $a_{1g}$ and $e_g^\pi$ Wannier functions are set to zero, i.e. the *pure* (or "un-hybridized") $a_{1g}$ and $e_g^\pi$ bands which correspond to the limit of the crystal-field splitting being much larger than the hybridization. These bands differ significantly from the LDA bands: the pure $e_g^\pi$ band, for instance, extends merely from $\min\epsilon_{e_g^\pi} \approx$ -0.5 eV to $\max\epsilon_{e_g^\pi} \approx$ 1.0 eV, and the pure $a_{1g}$ band, extending from $\min\epsilon_{a_{1g}} \approx$ -1.0 eV to $\max\epsilon_{a_{1g}} \approx$ 1.5 eV, tails off rather than peaks at the bottom of the band. These observations show that the peak at the bottom of the $a_{1g}$ DOS is due to $e_g^\pi$-hybridization rather than an bonding-antibonding splitting of the $a_{1g}$ levels. We note in passing that these $e_g^\pi$ and $a_{1g}$ bands, with the latter displaced upwards by $\sim$ 2 eV, are similar to the majority $e_g^\pi$ and $a_{1g}$ bands resulting from an LDA+U calculation for a ferromagnetically ordered approximation to the AFI phase[17,18].

### 2. *Effects of the imaginary part of the self-energy: orbital-selective coherence*

To see how strong Coulomb interactions alter the low energy properties of the system, we calculate the self-energy, which encodes all correlations beyond the LDA. We briefly recall what is expected on general grounds from Fermi liquid theory, before proceeding to a detailed analysis of the orbital-resolved self-energy $\Sigma_m(\omega)$ as calculated from LDA+DMFT.

If, at low energies, the imaginary part of the self-energy is small, well-defined quasi-particles exist and their dispersion $\omega_j(\mathbf{k})$ is given by the poles of the $\mathbf{k}$-resolved Green's function:

$$\det\left[\hat{H}^{LDA}(\mathbf{k}) + \Re\hat{\Sigma}(\omega+i0^+) - \mu - \omega\right] = 0 \quad (9)$$

In a Fermi liquid, sharp quasi-particles are expected at low energies and temperatures below the coherence temperature. In fact, in that regime the imaginary part of the self-energy behaves as

$$\Im\Sigma_m(\omega) = -B_m\left[\omega^2 + (\pi T)^2\right] + O\left(\omega^4\right),$$

thus indicating the existence of infinite-lifetime excitations at zero temperature and on the Fermi surface. However, quasi-particle lifetimes become shorter when moving away from the Fermi surface or at finite temperature, and fade away completely beyond the coherence temperature.

In $V_2O_3$, low-temperature properties are dominated by the onset of antiferromagnetic order. The paramagnetic phase of the pure zero-pressure compound only exists down to about 150 K. For technical reasons related to the high computational cost of QMC calculations at low temperatures, our calculations are done at an even higher temperature. The first issue we have to address in the analysis of the self-energies is thus the question of coherence.

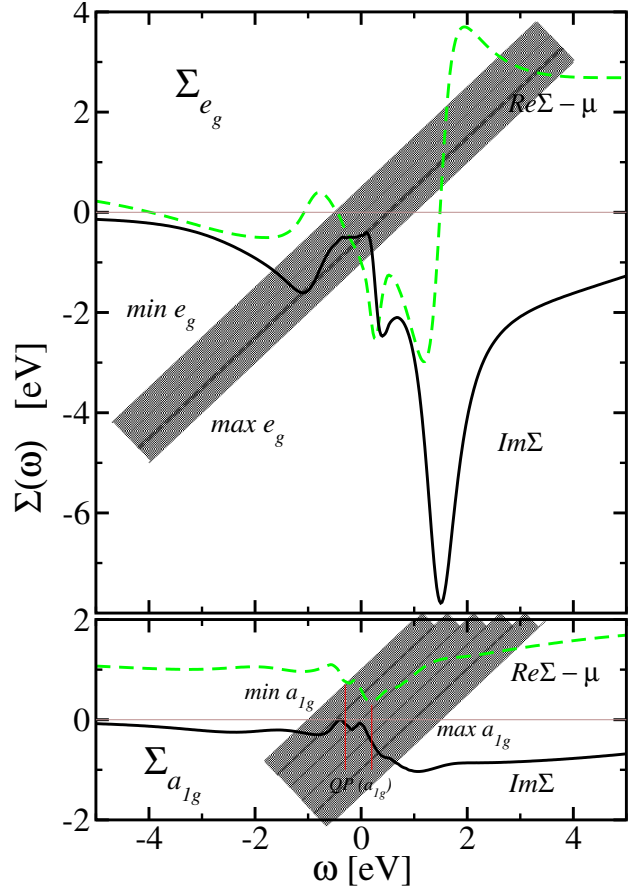

FIG. 1: (Color online) $e_g^\pi$ and $a_{1g}$ self-energies, $\Sigma_m(\omega)$, as functions of real energy $\omega$, calculated by LDA+DMFT using $U$=4.2 eV, $J$=0.7 eV, and $T$=390 K. Solid (black) and dashed (green) lines are respectively the imaginary and real parts of the self-energies. The chemical potential $\mu$=-3.63 eV was subtracted from the real parts. The grey stripes show the extent of the unhybridized LDA bands. Their intersection with the real part of the self-energies roughly indicate solutions of Eq. (9). These solutions give rise to coherent quasi-particles provided the imaginary part of the self-energy is small, which for the $a_{1g}$ orbitals is true within an energy range around the Fermi level indicated by the vertical red lines. For the $e_g^\pi$ the large imaginary part indicates that these orbitals stay in an incoherent regime all the way down to 390 K. See text for a discussion.

Fig. 1 displays the real and imaginary part of both orbital components of the self-energy on the real frequency axis as obtained from our LDA+DMFT calculation at 390 K. As seen from the lower panel, the imaginary part of the self-energy for the $a_{1g}$ orbitals indeed exhibits a dip at the Fermi level with a small value of $\Im\Sigma(0)$. The

$e_g^\pi$ orbitals, however, display a zero-frequency value of the imaginary part of the self-energy as large as 0.45 eV, demonstrating that the coherence temperature of the $e_g^\pi$ orbitals has not yet been reached.

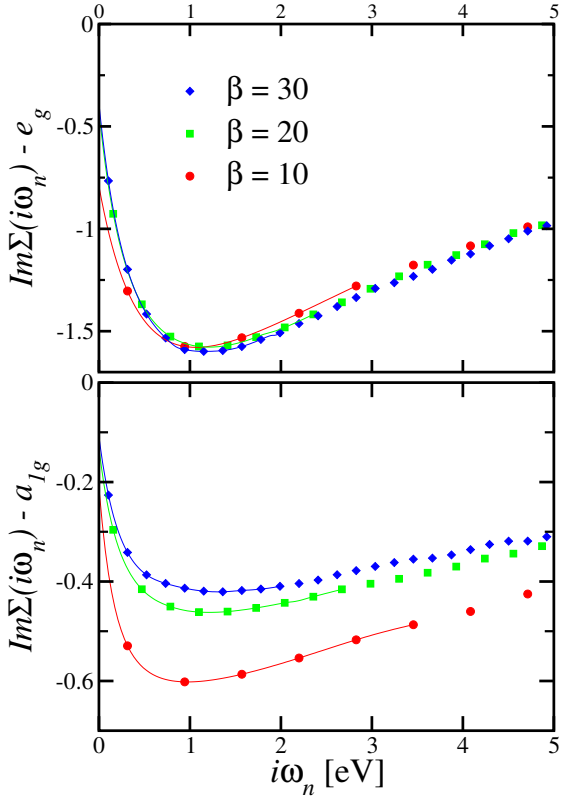

FIG. 2: (Color online) Imaginary parts of the Matsubara frequency self-energies for $e_g^\pi$ (top) and $a_{1g}$ (bottom) orbitals for different temperatures. (Red) dots, (green) squares and (blue) diamonds correspond to $\sim$ 1160 K, 580 K and 390 K, respectively. Symbols represent raw data while lines are extrapolations to zero frequency. $U$=4.2 eV, $J$=0.7 eV.

In Fig. 2 we show the imaginary parts of the self-energies on the imaginary axis, which are a direct output of the QMC. Calculations were performed for $T \sim$ 390, 580 and 1160 K ($\beta = 30, 20$ and 10 eV$^{-1}$) and allow to study the evolution of the self-energies with temperature. We first note that the zero-frequency value for the lowest temperature coincides with the ones of the real-axis self-energies displayed in Fig. 1, thus confirming the observation of the $a_{1g}$ orbitals being much closer to their coherent regime than the $e_g^\pi$ orbitals. The changes with temperature of the self-energy of the scarcely occupied $a_{1g}$ orbital is huge, possibly due to temperature-induced changes in the occupation, which are large on a relative scale. Indeed, the occupation of the $a_{1g}$ orbital evolves from 0.31 via 0.23 to 0.20 electrons per V atom for $T$ decreasing from 1160 K, over 580 K, to 390 K ($\beta$=10, 20, 30 eV$^{-1}$). The radical change between the 1160 K and 580 K curves may correspond to reaching the coherent regime for the $a_{1g}$ orbitals. For the $e_g^\pi$ orbitals, however, the temperature dependence seems to be weaker, and they clearly stay incoherent all the way down to 390 K.

### 3. Effects of the real parts of the self-energy

We now comment on the shape of the self-energy curves on the real frequency axis, displayed in Fig. 1 for both orbital components, starting from low energy and moving towards higher energy. The first striking observation is that the zero-frequency values of the self-energies lead to a considerable enhancement of the effective crystal-field splitting:

$$\Delta_{\text{eff}} = \Delta_{\text{bare}} + \Re\Sigma_{a_{1g}}(0) - \Re\Sigma_{e_g^\pi}(0) \quad (10)$$

Here, $\Delta_{\text{bare}}$ is the crystal field splitting of the unhybridized ("pure") LDA bands (see below). While $\Delta_{\text{bare}}$, as mentioned above, is of the order of 0.27 eV, the self-energy effects yield $\Delta_{\text{eff}} \simeq 1.9$ eV, hence a correlation-induced enhancement by a factor $\sim 7$, almost an order of magnitude ! The $a_{1g}$ orbital is pushed upwards in energy, while the $e_g^\pi$ orbitals are pushed downwards to an extent the that two bands hardly overlap anymore. Hence, this is accompanied by a strong enhancement of the orbital polarization in favor of the $e_g^\pi$'s. Because the occupancy of the $a_{1g}$ orbital is low, the frequency-dependence of the self-energy is significantly smaller ($\lesssim 1$ eV) for this orbital. Since the $a_{1g}$ and $e_g^\pi$ bands hybridize for each $\mathbf{k}$-point, increasing the crystal-field splitting changes the band structure and the constant-energy surfaces in a non-trivial way, which is explained in detail in the following.

When moving away from $\omega = 0$, the imaginary part of the $a_{1g}$ self-energy is small ($\propto \omega^2$) over a narrow range of frequency, from about -0.5 eV below to about 0.1 eV above the chemical potential. In that (Fermi-liquid) range, the real parts of the self-energy roughly obey a linear behavior[55]:

$$\Re\Sigma_m(\omega) = \Sigma_m(0) + (1 - 1/Z_m)\omega + O(\omega^3) \quad (11)$$

The $Z$'s are the quasi-particle spectral weights, corresponding also to the inverse of the effective mass enhancements. We find $m^*/m = 1/Z_{a_{1g}} \simeq 2.5$, indicating strong correlations; for the $e_g^\pi$ orbitals we formally calculate the weights to $1/Z_{e_g} \simeq 5$, even if – as discussed above – these orbitals are far from their Fermi liquid regime.

Let us finally turn to the high-frequency behavior of the $e_g^\pi$ self-energy (as mentioned, the $a_{1g}$ has a much weaker frequency dependence). For 2 eV < $|\omega|$ < 5 eV, $\Re\Sigma_{e_g}$ behaves approximately as $C_\pm + \Omega_\pm^2/\omega$, where $C_\pm$ and $\Omega_\pm$ control the shape and location of the Hubbard bands (see Fig. 1). For much larger $|\omega|$ (outside the range of the Fig. 1) we checked that

$$\Sigma(\pm\infty)_m = \Sigma_m^{Hartree} = Un_m + (2U - 5J)\sum_{m' \neq m} n_{m'}$$

is equal to the Hartree term (≈4.6 eV and 4.8 eV for the $e_g^\pi$ and $a_{1g}$ orbitals, respectively). In order to match together the low-frequency and high-frequency behavior, the $e_g^\pi$ self-energy must display a very strong frequency dependence at intermediate frequencies, e.g. for $\omega \simeq +1.7$ eV on the positive frequency side. In this region, the self-energy is well approximated by $\Sigma(+\infty)+\frac{\Omega_+^2}{\omega-1.7\text{eV}}$, corresponding to an almost pole-like behavior. Correspondingly, the imaginary part displays a pronounced peak at $\omega \simeq 1.7$ eV whose main effect is to strongly deplete the spectral weight in this energy range (see Par. III A 5 below and Fig. 4).

### 4. Correlated band structure

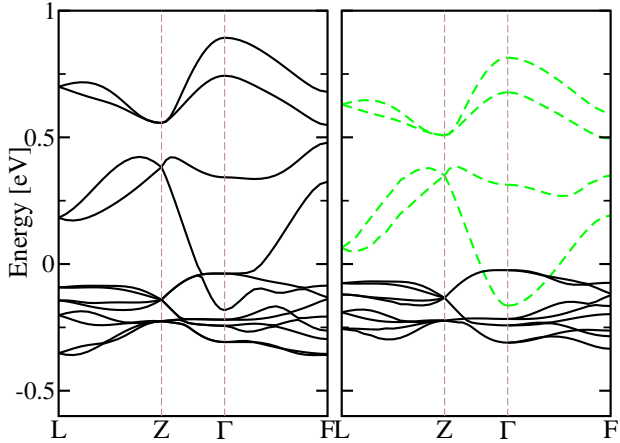

FIG. 3: (Color online) LDA+DMFT quasi-particle bands obtained by truncating the self-energies after linear order in $\omega$. The left-hand panel gives the eigenvalues of the re-normalized Hamiltonian (12) and the right-hand side gives the pure $e_g^\pi$ (black) and $a_{1g}$ (dashed green) band structure, scaled individually according to the scalar expression (13). The self-energy parameters $Z_m$ and $\Sigma_m(0)$ are given in Table I. Note that for the $e_g^\pi$ orbitals and for the $a_{1g}$ orbitals beyond $\sim 0.5$ eV the linearized bands do not correspond to true quasi-particle states due to a sizeable imaginary part of the self-energy (see discussion in the text). Our standard values, $U=4.2$ eV, $J=0.7$ eV and $T \sim 390$ K, were used.

As seen above, the imaginary part of the self-energy of the $a_{1g}$ orbitals is tiny in a small energy range around the Fermi level. This motivates us to attempt a description in terms of effective bands, even if, as seen above, one should not think in terms of quasi-particle excitations for the $e_g^\pi$ orbitals. First, we shall limit ourselves to a simplified but instructive graphical discussion. In the case of degenerate bands we find the solutions of Eq. (9) for a given energy $\epsilon$ as the intersections $\Re\Sigma(\omega) = \omega + \mu - \epsilon$. In other words, the solutions lie where the self-energy crosses a stripe that is linear in frequency and which extends over the LDA bandwidth over which $\epsilon$ varies. In the non-degenerate case a similar construction applies approximately when intersecting the real part of the self-energy with the unhybridized LDA band structure: In Fig. 1 we plot a stripe the width of which corresponds to the width of this unhybridized band structure[57].

In view of Fig. 1 and the above discussion we expect solutions over a wide range of frequency, extending roughly from -2 eV to 4 eV. However, only in regions where the imaginary parts of the self-energies are small, these will give rise to coherent quasi-particles, which is true within ±0.5 eV around the Fermi level, at least for the $a_{1g}$ orbital. The intersections in the range 2.5 to 4 eV correspond to the upper Hubbard band (see Fig. 1 and Fig. 4). In the low energy region where the imaginary part of the self-energy is small and can thus be neglected, we linearize the real part according to Eq. (11). The quasi-particle band structure can then be obtained from the eigenvalues of the matrix :

$$Z_m^{1/2} \left[ H_{Im,I'm'}^{\text{LDA}}(\mathbf{k}) + (\Re\Sigma_m(0) - \mu)\delta_{mm'}\delta_{II'} \right] Z_{m'}^{1/2} \quad (12)$$

The corresponding (linearized) band structure is plotted in the left panel of Fig. 3. The downward shift of the $e_g^\pi$ bands and the upwards shift of the $a_{1g}$, corresponding to the strong correlation-induced enhancement of the effective crystal field, is clearly visible on this figure. Moreover we expect that this enhanced crystal field leads to a strong decrease of the hybridization between the $a_{1g}$ and $e_g^\pi$ bands. Indeed, a simple calculation suggests that the local orbital character of the band is controlled by the parameter $V_\mathbf{k}/\Delta_{\text{eff}}$, with $V_\mathbf{k}$ the bare hybridization (see below). Therefore, we would expect that the hybridization diminishes to a point where the band shapes are close to those of the "pure" (i.e. unhybridized) bands in the lower, left-hand side of Fig. I.9. The right panel in Fig. 3 demonstrates that this expectation is indeed correct. In this figure, we display the quantities:

$$Z_m \left[ \epsilon_{mj}(\mathbf{k}) - \mu + \Re\Sigma_m(0) \right] \sim \omega_{mj}(\mathbf{k}), \quad (13)$$

that is, the unhybridized LDA bands $\epsilon_{mj}(\mathbf{k})$ each shifted and re-normalized according to its linearized $\Re\Sigma_m(\omega)$. The good agreement between the two panels in Fig. 3 demonstrates that the low energy quasi-particle band structure can be entirely understood as a combination of (i) a considerable enhancement of the effective crystal field and a corresponding un-hybridization of the $a_{1g}$ and $e_g^\pi$ orbitals and (ii) a (Brinkman-Rice) bandwidth reduction given by the $Z_m$'s, i.e. a mass enhancement of about 2.5 for the $a_{1g}$ and 5 for the $e_g^\pi$ orbitals. This is entirely consistent with Sawatzky's line of argument[17,23]. As a result, the $a_{1g}$ band *nearly empties* and its bottom tail merely straddles off the top of the $e_g^\pi$ band, similar to what was found in LDA+U calculations for the AFI structure[17,18]. Correspondingly, the Fermi surface is strongly modified by the correlations.

At low energy, the frequency-dependent self-energies, $\Sigma_m(\omega)$, shift, shape, and re-hybridize the bands in a way that reminds the effect of the potential functions in canonical band theory, in particular for alloys in the coherent potential approximation[46].

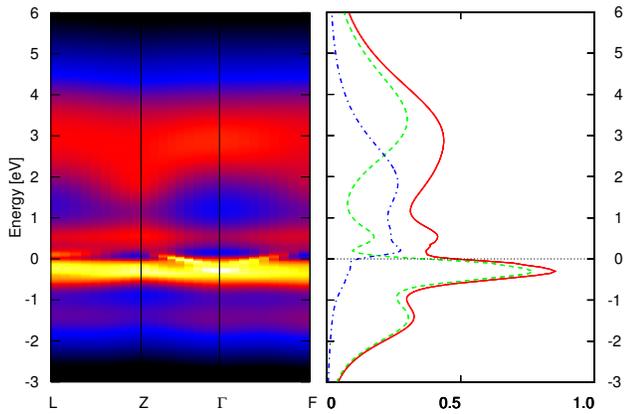

FIG. 4: (Color online) Left-hand side: Intensity plot of the **k**-resolved spectral function $A(\mathbf{k},\omega)$ calculated by LDA+DMFT for $U$=4.2 eV, $J$=0.7 eV and $T \sim$390 K. The total and orbitally resolved spectral functions are shown on the right-hand side. Solid (red) line represents the total spectral function, dashed (green) and dot-dashed (blue) lines are for the $e_g^\pi$ and $a_{1g}$ orbitals respectively.

5. *Momentum-resolved spectral function*

When leaving the low-frequency region in which the self-energies may be linearized, the full complex self-energies must be used and the spectral function $A(\mathbf{k},\omega) = -\frac{1}{\pi}\text{Tr}\Im G(\mathbf{k},\omega)$ has to be calculated. The result is given in Fig. 4 for our favorite set of parameters, $U$=4.2 eV, $J$=0.7 eV and $T \sim$ 390 K. The quasi-particle bands discussed above are clearly visible on this intensity plot, and fairly well defined. They lie in the region from -0.5 to 1 eV. Only the top of the $a_{1g}$ band becomes diffuse, and this, only away from the zone center, $\Gamma$. The nearly half-filled $e_g^\pi$ band transfers weight to incoherent Hubbard bands: a weak, narrow, lower band near -1.5 eV and a strong, broad, upper Hubbard band centered at $U'$ and extending from 2 to 4 eV. The depletion of the spectral weight in the $\sim$ +1.7 eV range is due to the large imaginary part of the self-energy in this energy range. The lower Hubbard band and the bottom of the upper Hubbard band disperse a bit. On the right-hand side of Fig. 4 we show the $a_{1g}$ and $e_g^\pi$ momentum-integrated (i.e on-site) spectral functions. The large, narrow $e_g^\pi$ peak $\left(1/Z_{e_g^\pi} = 5\right)$ slightly below the Fermi level, and the broader $a_{1g}$ quasi-particle peak $\left(1/Z_{a_{1g}} = 2.5\right)$ above the Fermi level are easily recognized. The remaining $e_g^\pi$ spectral weight is transferred to higher energies, where Hubbard bands, reminiscent of atomic-like excitations, are developed. Their presence already in the PM state is a characteristic feature of a correlated metal.

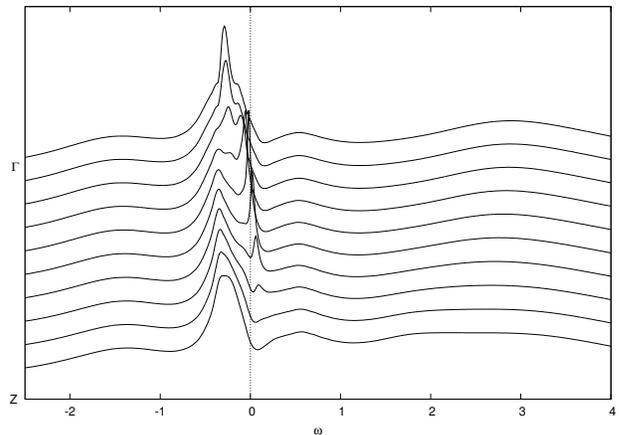

FIG. 5: Momentum-resolved spectra of the PM phase of $V_2O_3$ along the $\Gamma$-Z direction[39]. Parameters as in Fig. 4

Finally, in Fig. 5 we plot the **k**-resolved spectra for the sake of comparison with future angle-resolved photoemission experiments.

B. **Comparison with photoemission experiments**

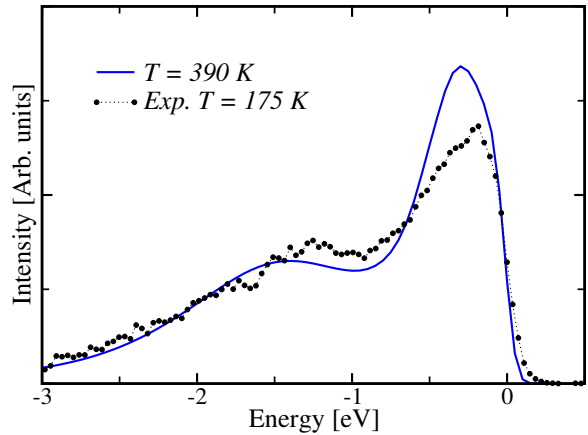

FIG. 6: (Color online) Comparison of the recent experimental PES at $T$=175 K of Mo *et al.*[8] (black dots) with the total spectral function (blue solid) from Fig. 4, convoluted with the Fermi function for $T$=175 K and broadened with the experimental resolution of 90 meV.

In Fig. 6 we compare the DMFT spectrum, multiplied with the Fermi function and broadened by an estimated experimental resolution of 90 meV, with the recent, more bulk-sensitive photoemission (PES) data of Mo *et al.*[8] obtained at $T$=175 K. Although for reasons of computational cost, our DMFT calculation of the spectral function was performed at 390 K, the agreement is excellent. The positions and relative weights of the quasi-particle peak and the weak, lower Hubbard peak, as well as the approximate form of the former, are correctly captured by the theoretical curve. That the height of the experimental quasi-particle peak is slightly lower than in theory, might be due to remaining surface contributions in the experiment, even in the spectra taken at a photon energy of 700 eV. Now that we have unveiled the nature of the quasi-particle peaks and the importance of re-hybridization effects due to correlations, the reason why our calculation agrees better with the PES data than the calculations of Keller et al.[21], whose quasi-particle peak was too narrow, is obvious. The Hamiltonian implementations of LDA+DMFT[24,29,43,44,48] such as the one presented in this paper thus seem to be important when aiming at a quantitative description of the physics of correlated materials.

### C. Temperature-dependence, coherence scales and local moments

The temperature dependent gap filling in the insulating phase has been studied extensively, both experimentally and theoretically in[9]. Here we focus on the temperature dependence of one-particle quantities and orbital resolved local susceptibilities in the high-temperature metallic phase. On general grounds, correlated metals are expected to display a crossover from coherent low-temperature behavior to bad-metal high-temperature properties, at a temperature corresponding to the quasi-particle coherence scale of the material. For $V_2O_3$ we have found this coherence scale to be orbital-selective and to be of the order of 400 K for the $a_{1g}$ orbitals, but substantially lower for the $e_g^\pi$. This order of magnitude is consistent with the sizeable mass enhancement[10] (and with the experimentally measured bad conductivity of the metallic phase[3]).

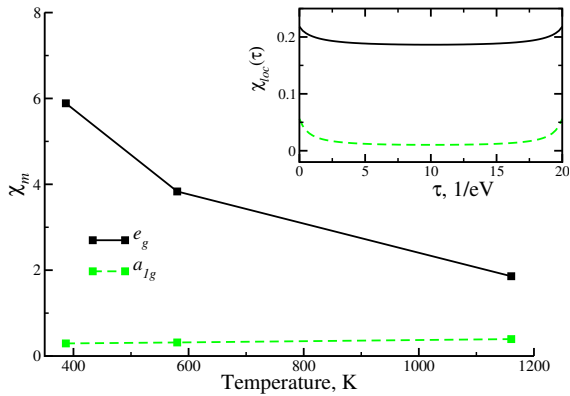

FIG. 7: (Color online) The orbital components of the local magnetic susceptibility $\chi_m$ versus temperature for $U$=4.2 and $J$=0.7 eV. Inset: $\chi_{loc}(\tau)$ for 580 K. (Green) dashed and (black) solid lines correspond to $a_{1g}$ and $e_g^\pi$ respectively.

We have also calculated the orbital-resolved local spin susceptibility $\chi_m = \int d\tau \chi_m^{loc}(\tau)$ from the local spin-spin correlation function $\chi_m^{loc}(\tau) = \frac{1}{4}\langle(n_{m\uparrow}(\tau) - n_{m\downarrow}(\tau))(n_{m\uparrow}(0) - n_{m\downarrow}(0))\rangle$. These quantities, plotted as a function of temperature in Fig. 7 display clear local moment behavior (corresponding to an increasing susceptibility upon lowering $T$) for the $e_g^\pi$ orbitals, while the $a_{1g}$ susceptibility is flat and small, corresponding to the tiny occupation of these orbitals and their more delocalized band-like character. For the $e_g^\pi$ orbitals, we further note that for temperatures between 600 and 400 K, one has not yet reached the regime of saturation where the local moment would be screened. This is in line with the above observation of an incoherent regime of the $e_g^\pi$ orbitals down to 390 K.

### D. Crystal-field enhancement for increasing $U$

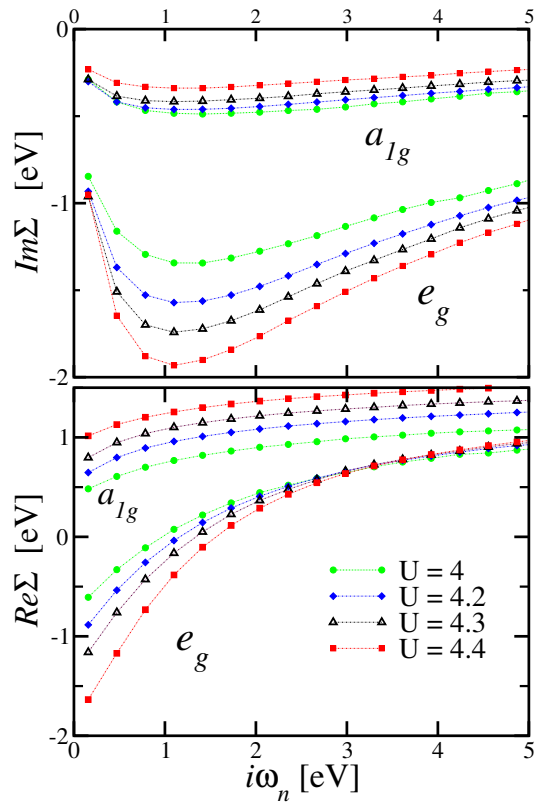

FIG. 8: (Color online) Imaginary and real parts of the self-energies, $\Sigma_m(i\omega_n)$, for the $e_g^\pi$ and $a_{1g}$ orbitals as functions of imaginary Matsubara-frequency and for different values of the Coulomb interaction $U$. (Green) dots, (blue) diamonds, (open black) triangles and (red) squares correspond to $U$=4, 4.2, 4.3 and 4.4 eV, respectively. For $U$=4.4 eV we find an insulating solution. $J$=0.7 eV, $T$=580 K.

We have seen above that one of the key effects of strong correlations for this material is to considerably enhance the effective crystal field splitting between the $e_g^\pi$ and $a_{1g}$ orbitals, and to correspondingly increase the orbital polarization in favor of the $e_g^\pi$. In this section, we briefly discuss how these effects evolve as the value of the parameter $U$ is increased.

| $U$ (eV) | | 0 | 4.0 | 4.2 | 4.3 | 4.4 | 4.2 |
|---|---|---|---|---|---|---|---|
| $T$ (K) | | 0 | 580 | 580 | 580 | 580 | 390 |
| $\Re\Sigma_m(0^+) - \mu$ | $e_g^\pi$ | 0 | -0.76 | -1.00 | -1.39 | -1.91 | -1.01 |
| (eV) | $a_{1g}$ | 0 | 0.42 | 0.56 | 0.69 | 0.93 | 0.59 |
| $\Delta_{eff}$ (eV) | | 0.27 | 1.45 | 1.83 | 2.35 | 3.11 | 1.87 |
| $1/Z_m$ | $e_g^\pi$ | 1 | 3.33 | 5 | 5 | ins. | 5 |
| | $a_{1g}$ | 1 | 2 | 2.5 | 2.5 | ins. | 2.5 |
| $n_m$ | $e_g^\pi$ | 1.48 | 1.70 | 1.78 | 1.82 | 1.88 | 1.80 |
| | $a_{1g}$ | 0.52 | 0.30 | 0.22 | 0.18 | 0.12 | 0.20 |

TABLE I: Effective crystal field splitting (Eq. 10), extrapolated to zero, real parts of the self-energy, quasi-particle parameters and Wannier-function occupations for increasing values of $U$ ($J$=0.7 eV is fixed). Note that the $Z$ factors for the $e_g^\pi$ orbitals are purely formal quantities, see text for a discussion. The values at $U$=0 eV are the LDA ones. The LDA $t_{2g}$ bandwidth is 2.5 eV while those of the pure $e_g^\pi$ and $a_{1g}$ LDA bands are respectively 1.5 and 2.5 eV. The last column with $U$=4.2 eV and smaller temperature $T \sim 390$ K corresponds to the figures discussed in Sections III A and III B. Note that $U$=4.2 eV is our favorite choice for the PM phase.

The quasi-particle weights, the Wannier-orbital populations and the re-normalized crystal-field are given in Table I for several values of $U$. These low-frequency quantities were extracted directly from the imaginary frequency QMC self-energies displayed in Fig. 8 (the $Z_m$'s are obtained by extrapolating the low-frequency behavior of the self-energy to zero frequency and analyzing the slope of the linear regime of $\Im\Sigma_m(i\omega_n)$). As discussed above, the $a_{1g}$ self-energy has a weaker frequency dependence, because of its lower occupancy, indicating smaller correlation effects. For both Wannier orbitals, the low-frequency spectral weight $Z_m$ decreases with increasing Coulomb interaction as expected. At the same time, $\Im\Sigma_{a_{1g}}(i\omega_n)$ becomes less frequency-dependent with increasing $U$, while for the strongly correlated $e_g^\pi$ orbital, the behavior of $\Im\Sigma_{e_g}(i\omega_n)$ is the opposite: it becomes more correlated with increasing $U$, corresponding to an increasing population of this orbital.

As seen from Table I, the low-frequency effective crystal field increases dramatically as $U$ is increased. The LDA bands in the left-hand side of Fig. I.10 yield a distance of $\max\epsilon_{e_g^\pi} - \min\epsilon_{a_{1g}} = 2.02$ eV from the top of the $e_g^\pi$ band to the bottom of the $a_{1g}$ band. Since both band extrema are at the $\Gamma$, point where $a_{1g}$ and $e_g$ symmetries cannot mix, the hybridized and the pure LDA bands have the same value at this point. The above value is roughly the self-energy shift needed to push the top of the $e_g^\pi$ band below and the bottom of the $a_{1g}$ band above the Fermi level and, hence, to cause a metal-insulator transition while keeping the structure of pure V$_2$O$_3$ at normal pressure.

From Table I we realize that the critical value $\Delta_{\text{eff},c} = 2.02 + 0.27 = 2.29$ eV is reached for $U_c = 4.3$ eV. When this value of the Coulomb repulsion is exceeded, the system becomes an insulator with a direct gap (at $\Gamma$) between a valence band of $e_g^\pi$ and a conduction band of $a_{1g}$ character.[58] Albeit driven by on-site Coulomb interactions, this transition is not caused by a divergence of the self-energy at low-frequency (like in the Mott transition of the single band model at particle-hole symmetry), but rather by the fact that effective crystal field splitting (or $\Sigma_{a_{1g}}(0) - \Sigma_{e_g^\pi}(0)$) is so large that the $e_g^\pi$ and $a_{1g}$ bands no longer overlap. The population of the $e_g^\pi$ band monotonously increases towards 2 and that of the $a_{1g}$ band decreases towards 0. This is fairly similar to what was recently found for the pressure-induced insulator-metal transition in LaMnO$_3$[47].

We emphasize that this study as a function of $U$ was performed by keeping the crystal structure of pure V$_2$O$_3$ at normal pressure. However, it is clear from this study that a slight decrease of the distance from the top of the $e_g^\pi$ to the bottom of the $a_{1g}$ LDA band suffices to drive the system from the metallic to the insulating phase. Indeed, the LDA bands for the expanded structure of 3.8% Cr-doped V$_2$O$_3$ given on the right-hand side of Fig. I.10 yield: $\max\epsilon_{e_g^\pi} - \min\epsilon_{a_{1g}} = 1.69$ eV, that is a 0.33 eV decrease. This large value is consistent with the fact that the observed conductivity gap[3] in the PI phase always exceeds 0.3 eV. Of this 0.33 eV reduction caused by the Cr-induced expansion of the lattice, 0.17 eV are due to the bottom of the $a_{1g}$ band moving up with respect to the $a_{1g}$ on-site energy, 0.13 eV are due to to the top of the $e_g^\pi$ band moving down, and only 0.03 eV are due to the increase of the crystal-field splitting. Further details are given in paper I. With this, we believe to have demystified the nature of the metal-insulator transition in V$_2$O$_3$.

### E. Hamiltonian DMFT versus fixed hybridization approximation

At this point, a word of comparison between the LDA+DMFT calculations performed in the Hamiltonian form (as in the present paper) and the ones using a density of states integration as an approximation, is in order. In practice, it often looks as if the error induced by doing the hybridization approximation (i.e. using a simple Hilbert transform of the density of states instead of the full $\mathbf{k}$-sum in the self-consistency condition (8)), could be partially compensated by a slightly bigger choice of the Coulomb interaction parameter $U$, see e.g.[53] in comparison with[29] or the discussion in[24]. In the present case of strong orbital and occupation shifts induced by the correlations, however, the effect of the hybridization approximation is particularly severe: as discussed above, one of the key effects of the correlations in V$_2$O$_3$ is the suppression of hybridization of the $a_{1g}$ and $e_g^\pi$ bands by the correlation enhanced crystal field splitting. The fixed hybridization approximation however neglects any correlation-driven changes of the hybridization, and thus entirely overlooks this effect.

This can be rationalized in a simple two-band model with a purely non-local hybridization $V_\mathbf{k}$, i.e. $V_\mathbf{k}$ such that $\sum_\mathbf{k} V_\mathbf{k}=0$ which leads to a diagonal self-energy $\Sigma$: The momentum-resolved Green's function within DMFT reads

$$G(\mathbf{k},\omega) = \left[\omega + \mu - \begin{pmatrix} \epsilon_\mathbf{k}^1 & V_\mathbf{k}^* \\ V_\mathbf{k} & \epsilon_\mathbf{k}^2 \end{pmatrix} - \begin{pmatrix} \Sigma_1 & 0 \\ 0 & \Sigma_2 \end{pmatrix}\right]^{-1} \quad (14)$$

$$= \frac{1}{(\omega + \mu - \epsilon_\mathbf{k}^1 - \Sigma_1)(\omega + \mu - \epsilon_\mathbf{k}^2 - \Sigma_2) - |V_\mathbf{k}|^2}$$

$$\times \begin{pmatrix} \omega + \mu - \epsilon_\mathbf{k}^2 - \Sigma_2 & V_\mathbf{k} \\ V_\mathbf{k}^* & \omega + \mu - \epsilon_\mathbf{k}^1 - \Sigma_1 \end{pmatrix}$$

Both, the diagonal and off-diagonal (hybridization-induced) components of the Green's function are thus re-normalized by the denominator containing the self-energy, and the corresponding eigenvectors adjust at each cycle of the DMFT self-consistency. The fixed hybridization approximation consists instead in replacing the Hamiltonian matrix by its diagonalized version while keeping the self-energy in the form above.

$$G(\mathbf{k},\omega) = \left[\omega + \mu - \begin{pmatrix} E_\mathbf{k}^{(+)} & 0 \\ 0 & E_\mathbf{k}^{(-)} \end{pmatrix} - \begin{pmatrix} \Sigma_+ & 0 \\ 0 & \Sigma_- \end{pmatrix}\right]^{-1}$$

where

$$E_\mathbf{k}^{(\pm)} = \frac{1}{2}(\epsilon_\mathbf{k}^1 + \epsilon_\mathbf{k}^2) \pm \frac{1}{2}\sqrt{(\epsilon_\mathbf{k}^1 - \epsilon_\mathbf{k}^2)^2 + 4|V_\mathbf{k}|^2}$$

denote the eigenvalues of the LDA Hamiltonian. Technically, this form is easier to handle because $\mathbf{k}$-sums can be taken by simply integrating over the LDA partial densities of states. There is, however, no way of updating the orbital character that in general will be modified under the influence of the correlations. In fact, assuming two bands split by a crystal field splitting $\Delta$ but otherwise degenerate (i.e. $\epsilon_\mathbf{k}^1 = \epsilon_\mathbf{k}^2 + \Delta$) the eigenvectors of the full matrix problem (14) depend on the self-energy via the re-normalized crystal field splitting $\Delta_{eff} = \Delta + \Sigma_1(0) - \Sigma_2(0)$. The eigenvectors read:

$$\cos(\theta)|1\rangle \pm \sin(\theta)|2\rangle$$

with

$$\tan(\theta) = \frac{\Delta_{eff}}{2V_\mathbf{k}} + \sqrt{\left(\frac{\Delta_{eff}}{2V_\mathbf{k}}\right)^2 + 1}$$

The orbital character thus depends on the ratio of the *re-normalized* crystal field splitting over the hybridization,

as announced above. In the fixed hybridization approximation, however, the re-normalized crystal field splitting $\Delta_{eff}$ in the above decomposition of the eigenvectors onto the basis of the non-interacting non-hybridized single-particle states $|1\rangle, |2\rangle$ is replaced by the bare crystal field splitting $\Delta$. Thus, the orbital characters are fixed once and for all by the LDA Hamiltonian and cannot respond to the correlations.

This approximation is particularly severe in the present case, since the strong enhancement of the crystal field splitting profoundly modifies the effective hybridization.

### F. The paramagnetic insulating phase

We have performed an LDA+DMFT calculation using the expanded crystal structure of the 3.8% Cr-doped material using the same value $U = 4.2$ eV as the one used in our study of pure $V_2O_3$. We note however that because the screening in the PI phase is expected to be somewhat less efficient than in the PM phase, it might have been appropriate to also consider a value of $U$ slightly larger. We nevertheless keep the same value for the sake of comparison.

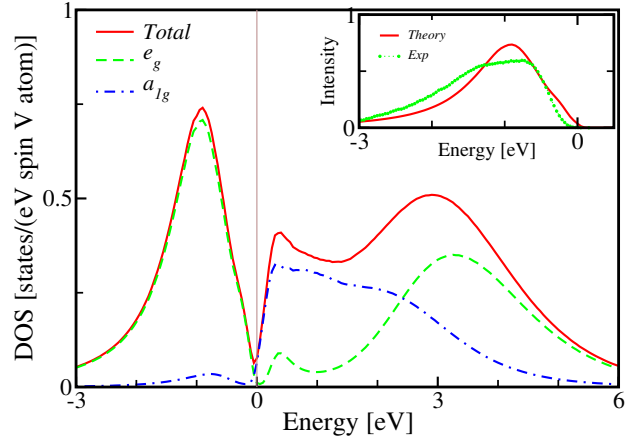

FIG. 9: (Color online) Spectral function for insulating $(V_{0.962}Cr_{0.038})_2O_3$ at $T=580$ K ($\beta=20$ eV$^{-1}$) calculated by DMFT using LDA $t_{2g}$ Wannier functions. Solid (red) line represents total spectral function, dashed (green) and dot-dashed (blue) lines represent $e_g^\pi$ and $a_{1g}$ orbitals respectively. Th inset shows a comparison of recent experimental PES at $T=175$ K of Mo *et al.*[9] (green dots) with the total spectral function convoluted with the Fermi function for $T=175$ K and broadened with the experimental resolution of 90 meV. $U=4.2$ eV and $J=0.7$ eV. The chemical potential is at zero energy.

The momentum-integrated, orbitally resolved, spectral functions resulting from this calculation (after a MaxEnt treatment of the data) are displayed in Fig. 9. It is seen that indeed the quasi-particle features at the Fermi level are almost entirely suppressed. Close examination of the figure (and of the quasi-particle bands - not shown) reveal that the low-frequency spectral weight is small but finite.

Finally, we comment on a technical aspect of this calculation of the PI phase. Because of the almost complete orbital polarization, it proved to be essential to allow for global updates in the QMC calculation. In contrast to the calculations in the PM phase which used the Hirsch-Fye algorithm, we have employed the continuous-time Rubtsov algorithm[54] in the PI phase, with global updates, in order to overcome the tendency of the HF algorithm with local updates to get trapped in a given fully polarized configuration.

### G. Comparison with polarization dependent X-ray absorption

From polarization-dependent X-ray-absorption (XAS) measurements analyzed by means of cluster calculations, Park et al. [13] concluded that the PI phase is a mixture of the two-electron configurations $(e_g^\pi, e_g^\pi)$ and $(e_g^\pi, a_{1g})$ in the ratio 3 : 2, and that the PM phase a 1 : 1 mixture. This means that the *ratio* of $e_g^\pi$ to $a_{1g}$ electrons *in the V 3p core region* is 4 : 1 in the PI phase and 3 : 1 in the PM phase. This ratio is between different "local" V $3d$ electrons with the same radial function and normalized the same way, specifically: the ratio between the $e_g^\pi$ and $a_{1g}$ occupations of partial waves normalized in the atomic (MT) sphere. As explained in paper I, this is the same as the ratio between the occupations in a large basis of nearly orthonormal LMTOs. The distribution of such characters is what is shown by the fat bands in Fig. I.4.

In paper I, we found that in the O $2p$ band, there are 0.32 local $e_g^\pi$ electrons and 0.18 local $a_{1g}$ electrons on each vanadium. The PI and PM structures gave the same numbers, which is consistent with the fact that the V-O distances are essentially the same.

Coming now to the V $t_{2g}$ bands, paper I explained that the $a_{1g}$ NMTO Wannier function has 0.60 local $a_{1g}$, and 0.11 local $e_g^\pi$ character (sum of $e_{g,1}^\pi$ and $e_{g,2}^\pi$), and that each of the $e_g^\pi$ NMTO Wannier functions has 0.60 local $e_g^\pi$ character (sum of $e_{g,1}^\pi$ and $e_{g,2}^\pi$) and 0.054 $a_{1g}$ character.

For the PI phase, our DMFT calculation showed that of the two electrons in the *full* $e_g^\pi$ band, 0.10 have $a_{1g}$ Wannier character. This number is the integral over the occupied part of the blue, dot-dashed $a_{1g}$ DOS in Fig. 9, and it is caused by the hybridization between the $a_{1g}$ Wannier orbital and the two $e_g^\pi$ Wannier orbitals. In paper I, this number was estimated roughly using second-order perturbation theory as the real-space sum $\sum t^2/\Delta_{\text{eff}}$, where $t$ are the $a_{1g}$-to-$e_g^\pi$ hopping integrals and $\Delta_{\text{eff}} = 1.87$ eV. Also there, the result was 0.10. We took $\Delta_{\text{eff}}$ to be the same for the PI and PM phases because, as explained above, $\Delta_{\text{eff}}$ only increases by 0.03 eV when going from the PM to PI phase.

The number of local $a_{1g}$ electrons in the PI phase is thus:

$$0.18 + 0.60 \times 0.10 + 0.054 \times 1.90 = 0.34$$

and the number of local $e_g^\pi$ electrons is:

$$0.32 + 0.60 \times 1.90 + 0.11 \times 0.10 = 1.47,$$

yielding the ratio 1.47 : 0.34 = 4.3, in good agreement with the XAS result: 4.

For the PM phase, our DMFT calculation (Table I and Fig. 3) gave 0.22 $a_{1g}$ Wannier electrons. These increase by 0.12 $a_{1g}$ electrons compared with the insulating phase is caused by the band overlap seen in Fig. 2, and is the occupation of the metallic $a_{1g}$ electron pocket. Hence, the number of local $a_{1g}$ electrons in the PM phase is:

$$0.18 + 0.60 \times 0.22 + 0.054 \times 1.78 = 0.41$$

and the number of local $e_g^\pi$ electrons is:

$$0.32 + 0.60 \times 1.78 + 0.11 \times 0.22 = 1.41,$$

yielding the ratio 1.41 : 0.41 = 3.4, again in good agreement with the XAS result: 3.

## IV. CONCLUSION

Summarizing, we have presented a detailed study of the paramagnetic metallic and insulating phases of $V_2O_3$ within an $NMTO$-$t_{2g}$-Hamiltonian implementation of LDA+DMFT. We have discussed the strong correlation-induced changes in the low energy quasi-particle band structure of the metallic phase. A correlation-induced enhancement of the crystal field splitting leads to a un-hybridization of the $a_{1g}$ and $e_g^\pi$ bands and a nearly empty $a_{1g}$ band. The resulting spectral functions are in excellent agreement with recent photoemission experiments. From a technical point of view, the drastic modifications that correlation effects cause in the hybridization of the orbitals, explain why LDA+DMFT calculations that neglect these changes compare less favorably to photoemission experiments than Hamiltonian calculations[20,21,24]. We have calculated local and momentum-resolved spectral functions, self-energies, quasi-particle weights and local spin susceptibilities and find PM $V_2O_3$ to be a rather strongly correlated metal with $Z$ of the order of 0.2 and a coherence temperature below 400 K. Interestingly, the material develops strongly orbital-selective coherence scales, with the $e_g^\pi$ orbitals exhibiting quasi-particle lifetimes shorter than $1/(0.5eV)$ on the Fermi level down to 390 K. Still, our picture is different from the one by Laad et al.[26] who describe the paramagnetic phase as an orbital-selective Mott phase with a gap (or – due to finite temperature – pseudogap) in the $e_g^\pi$ orbitals.

We have discussed the transition from the paramagnetic metallic to the paramagnetic insulating phase, the origin of which we tracked down to the correlation-induced enhancement of the crystal field splitting. Finally, we have calculated the spectral function in the insulating phase, in good agreement with photoemission data. A technical advance employed for this latter calculation was the use of the continuous-time Quantum

Monte Carlo algorithm[54], which to our knowledge has not been applied to LDA+DMFT calculations before.

## V. ACKNOWLEDGMENT

We are grateful to J. Allen and S.-K. Mo for providing us with their experimental data. We acknowledge useful discussions with V. Anisimov, K. Held, G. Keller, G. Panaccione and D. Vollhardt. AIP acknowledges the Marie Curie IIF grant 021820. OKA and TSD acknowledge MPG-India partner group for the collaboration and G. Sawatzky for useful discussions at the initial stage of this work. ANR is grateful to Dynasty foundation and NWO (grant 047.016.005) for support. The hospitality of KITP (UCSB), where this work was started, is also acknowledged. This work was supported by a supercomputing grant at IDRIS Orsay under project number 061393.

## APPENDIX A: MULTI ORBITAL HIRSCH FYE ALGORITHM

As discussed above solving the DMFT equation requires – at each iteration – the solution of an effective local problem, that is the calculation of the Green's function (4) with the effective action (3). Using the fermionic path integral formalism, the problem can be stated as follows: for a given Weiss field, one needs to calculate the Green's function

$$G^\sigma_{mm'}(\tau-\tau') = -\frac{1}{Z}\mathcal{T}\int D[c^+_{m\sigma},c_{m'\sigma}]e^{-S_{eff}}c_{m\sigma}(\tau)c^+_{m'\sigma}(\tau'). \quad (A1)$$

where $\mathcal{T}$ denotes the time-ordering operator and Z the partition function

$$Z = \int D[c^+_{m\sigma},c_{m'\sigma}]e^{-S_{eff}} \quad (A2)$$

For the solution of this problem the auxiliary fields quantum Monte Carlo (QMC) method by Hirsch and Fye is used[49]. Here, we review this approach in the multi-orbital case.

The imaginary time is discretized in $L$ slices $\tau_i, i = 1, 2, ..., L$ of size $\Delta\tau = \beta/L$. Performing a standard Trotter decomposition one obtains for the partition function:

$$\begin{aligned} Z &= Tr\prod_{l=1}^L e^{-\Delta\tau(H^{LDA}+H^{int})} \\ &\simeq Tr\prod_{l=1}^L e^{-\Delta\tau H^{LDA}}e^{-\Delta\tau H^{int}} \end{aligned} \quad (A3)$$

The quadratic interaction term in the effective action (3) can be decoupled by discrete Hubbard-Stratonovich transformation introduced by Hirsch[50]

$$e^{-\Delta\tau U_{mm'}[n_m n_{m'}-(n_m+n_{m'})/2]} = \frac{1}{2}\sum_{S_{mm'}=\pm 1} e^{\lambda_{mm'} S_{mm'}(n_m-n_{m'})} \quad (A4)$$

where $S_{mm'}(\tau)$ are auxiliary Ising fields. Starting form here and in the following $m$ is a spin-orbital index and $U$ matrix is a general spin and orbital dependent interaction matrix. One field has to be introduced for each pair of spin-orbitals and time slice. The coupling of the auxiliary spins to the physical occupations is related to the original Coulomb interaction via

$$\lambda_{mm'} = \operatorname{arccosh}[e^{\frac{1}{2}\Delta\tau U_{mm'}}]. \quad (A5)$$

Using the Hirsch transformation one can integrate out the fermionic fields in the path integral. The resulting Green's and partition function matrices have the following form:

$$\hat{G}^\sigma = \frac{1}{Z}\frac{1}{2^{N_f L}}\sum_{\{S_{mm'}(\tau)\}=\pm 1} \hat{G}^\sigma(S_{mm'})W(S_{mm'}) \quad (A6)$$

$$Z = \frac{1}{2^{N_f L}}\sum_{\{S_{mm'}(\tau)\}=\pm 1} W(S_{mm'}(\tau)) \quad (A7)$$

with

$$W(S_{mm'}) \equiv \det[\hat{G}^{-1}_\uparrow(S_{mm'})]\det[\hat{G}^{-1}_\downarrow(S_{mm'})] \quad (A8)$$

where $N_f = M(2M-1)$ is the number of Ising fields for the $M$ interacting orbitals and $\hat{G}(S_{mm'})$ is the Green's function for a given configuration of the auxiliary Ising fields:

$$\begin{aligned} G^{-1}_{mm'}(\tau,\tau';S) &= \mathcal{G}^{-1}_{mm'}(\tau,\tau') + \Delta_m(\tau)\delta_{mm'}\delta_{\tau\tau'} \\ \Delta_m(\tau) &= e^{V_m(\tau)} - 1 \\ V_m(\tau) &= \sum_{m'}\lambda_{mm'} S_{mm'}(\tau)\sigma_{mm'} \end{aligned} \quad (A9)$$

Here we have introduced the generalized Pauli matrix:

$$\sigma_{mm'} = \begin{cases} +1, & m < m' \\ 0, & m = m' \\ -1, & m > m' \end{cases} \quad (A10)$$

In the multi band case at low filling a numerical problem with $\mathcal{G}$ related to the large absolute value of the chemical potential can occur. In this case a large part of the $\mathcal{G}(\tau)$ function can be very close to zero which can be the source of an inversion problem in eq. (A9). In order to avoid this problem a technical trick can be introduced: the bath Green's function and effective Hirsch potential are replaced by $\mathcal{G}^{-1} \to \mathcal{G}^{-1} - \varsigma$ and $V_m(\tau) \to V_m(\tau) + \varsigma\Delta\tau$. This replacement stems from a specific choice of dividing the non-interacting from the interacting parts of the Hamiltonian. It is exact for all values of the shift $\varsigma$, which can thus be chosen for numerical convenience.

In order to calculate the Green's function we need in principle to integrate over all possible Ising spin configurations $\{S_{mm'}(\tau)\}$. It is this sum that is performed by a Monte Carlo sampling procedure. The Green functions of two arbitrary configuration are related to each other by the following Dyson equation

$$G' = \left[1 + (1-G)\left(e^{V'-V} - 1\right)\right]^{-1} G. \quad (A11)$$

Calculating $G'$ from $G$ via this equation reduces the number of operations from the $(ML)^3$ operations necessary for a standard matrix inversion to $(ML)^2$.

Using QMC importance sampling we integrate over the Ising fields with the absolute value of $W(S_{mm'})$ as a stochastic weight. The Metropolis algorithm is used to perform QMC spin-flips in the space of the Ising fields. For a single spin-flip $S_{mm'}$ the determinant ratio is calculated as follows:

$$\det[\hat{G}]/\det[\hat{G}'] = R_m R_{m'} - R_{mm'} \quad (A12)$$

with

$$\begin{aligned} R_m &= 1 + [1 - G_{mm}(\tau,\tau)]\Delta_m(\tau) \\ R_{m'} &= 1 + [1 - G_{m'm'}(\tau,\tau)]\Delta_{m'}(\tau) \\ R_{mm'} &= G_{mm'}(\tau,\tau)\Delta_{m'}(\tau) G_{m'm}(\tau,\tau)\Delta_m(\tau) \end{aligned}$$

and the Green's function matrix is updated in the standard two-step update:

$$G'_{m_1 m_2}(\tau_1, \tau_2) = G_{m_1 m_2}(\tau_1, \tau_2) + \\ \left[G_{m_1 m}(\tau_1, \tau) - \delta_{m_1 m}\delta_{\tau_1,\tau}\right] \frac{\Delta_m(\tau)}{R_m(\tau)} G_{mm_2}(\tau,\tau_2) \quad (A13)$$

$$G^{new}_{m_1 m_2}(\tau_1, \tau_2) = G'_{m_1 m_2}(\tau_1, \tau_2) + \\ \left[G'_{m_1 m'}(\tau_1, \tau) - \delta_{m_1 m'}\delta_{\tau_1,\tau}\right] \frac{\Delta_{m'}(\tau)}{R_{m'}}(\tau) G'_{m'm_2}(\tau,\tau_2) \quad (A14)$$

The above described QMC algorithm allows one to calculate the Green's function in the general matrix form. This has several advantages: *(i)* it can be applied to a system with low crystal symmetry where the effective Weiss field and Green's function matrices cannot be presented in diagonal form, *(ii)* one can easily think about $m$ as site-spin-orbital index and thus without changing any single line in the QMC part of the program it can be used for cluster DMFT calculations for systems where e.g. pairs of atoms play an important role[51,52].

### APPENDIX B: CONTINUOUS-TIME QUANTUM MONTE CARLO

The continuous-time fermionic Quantum Monte Carlo technique differs from the Hirsch-Fye algorithm in several important aspects. It does not require time discretization and does not use any auxiliary fields. Consequently, the computational complexity is different. A particular advantage over the Hirsch-Fye method can be achieved for an interaction, which is non-local in space, orbital and spin indices, or in time. In this Appendix, we describe the general ideas of the method and the usage of so-called global updates, that are a way to get rid of local trappings of the CT-QMC random walker.

We start from the partition function $Z = \text{Tr}\mathcal{T} e^{-S}$ for a system with pair interaction in the most general case and split the action $S$ into two parts: an unperturbed action $S_0$ of Gaussian form and an interaction part $W$.

$$\begin{aligned} S &= S_0 + W, & (B1) \\ S_0 &= \int\int t_r^{r'} c_{r'}^\dagger c^r dr dr' \\ W &= \int\int\int\int w_{r_1 r_2}^{r'_1 r'_2} c_{r'_1}^\dagger c^{r_1} c_{r'_2}^\dagger c^{r_2} dr_1 dr'_1 dr_2 dr'_2 \end{aligned}$$

Here, $\mathcal{T}$ is the time-ordering operator, $r = \{\tau, s, i\}$ is a combination of the continuous imaginary-time variable $\tau$, spin projection $s$ and the discrete index $i$ numbering the single-particle orbital states. Integration over $dr$ implies the integral over $\tau$, and the sum over all lattice states and spin projections: $\int dr \equiv \sum_i \sum_s \int_0^\beta d\tau$.

Now we switch to the interaction representation and make a perturbation series expansion for the partition function $Z$ assuming $S_0$ as the unperturbed action:

$$\begin{aligned} Z &= \sum_{k=0}^\infty \int dr_1 \int dr'_1 ... \int dr_{2k} \int dr'_{2k} \Omega_k(r_1,...,r'_{2k}), \\ \Omega_k &= Z_0 \frac{(-1)^k}{k!} w_{r_1 r_2}^{r'_1 r'_2}...w_{r_{2k-1} r_{2k}}^{r'_{2k-1} r'_{2k}8} D_{r'_1 r'_2...r'_{2k}}^{r_1 r_2...r_{2k}}, \\ D_{r'_1...r'_{2k}}^{r_1...r_{2k}} &= <\mathcal{T} c_{r'_1}^\dagger c^{r_1}...c_{8r'_{2k}}^\dagger c^{r_{2k}} >_0 . \end{aligned} \quad (B2)$$

Here $Z_0 = \text{Tr}(\mathcal{T} e^{-S_0})$ is the partition function for the unperturbed system, the triangular brackets with the subscript "zero" denote the average over the unperturbed system: $<A>_0 = Z_0^{-1}\text{Tr}(\mathcal{T} A e^{-S_0})$. Since $S_0$ is Gaussian, one can apply Wick's theorem, and therefore $D_{r'_1...r'_{2k}}^{r_1...r_{2k}}$ is the determinant of a $2k \times 2k$ matrix which consists of the bare two-point Green's functions $g_{r'}^r = <\mathcal{T} c_{r'}^\dagger c^r >_0$:

$$D_{(2k)} \equiv D_{r'_1 r'_2...r'_{2k}}^{r_1 r_2...r_{2k}} = \det |g_{r_j}^{r_i}| \quad (B3)$$

An important property of the above formula is that the integrands remain unchanged under the permutations $r_i, r_{i'}, r_{i+1}, r_{i'+1} \leftrightarrow r_j, r_{j'}, r_{j+1}, r_{j'+1}$ with any $i, j$. Therefore it is possible to introduce a quantity $K$, which we call "state of the system" and that is a combination of the perturbation order $k$ and an *unnumbered set* of $k$ tetrades of coordinates. Now, denote $\Omega_K = k!\Omega_k$, where the factor $k!$ reflects all possible permutations of the arguments. In the series (B2) for the partition function, the $k!$ in the nominator and the denominator cancel each other, so that this series can be rewritten in the simple form

$$Z = \int \Omega_K D[K], \qquad (B4)$$

where $\int D[K]$ means the summation over $k$ and integration over all possible realizations of the above-mentioned unnumbered set at each $k$.

In a similar way, we can express the interacting two-point Green's function for the system (B1) using the perturbation series expansion (B2). It reads:

$$G_{r'}^{r} \equiv Z^{-1} < \mathcal{T} c_{r'}^{\dagger} c^{r} e^{-W} >_0 = Z^{-1} \int G_K \Omega_K D[K] \qquad (B5)$$

where $G_K(r, r')$ is defined as

$$G_K(r, r') = \frac{< \mathcal{T} c_{r'}^{\dagger} c^{r} c_{r_1'}^{\dagger} c^{r_1} ... c_{r_{2k}'}^{\dagger} c^{r_{2k}} >_0}{< \mathcal{T} c_{r_1'}^{\dagger} c^{r_1} ... c_{r_{2k}'}^{\dagger} c^{r_{2k}} >_0}$$

This is nothing else than the ratio of two determinants: $D_{(2k+1)}/D_{(2k)}$.

The important thing to notice is that the series expansion for an exponential *always* converges for finite fermionic systems. A mathematically rigorous proof can be constructed for Hamiltonian systems. Indeed, the many-body fermionic Hamiltonians $H_0$ and $W$ have a finite number of eigenstates that is $2^{N_{orb}}$, where $N_{orb}$ is the total number of electronic orbitals in the system. Now one can observe that $\Omega_k < const \cdot W_{max}^k$, where $W_{max}$ is the eigenvalue of $W$ with maximal modulus. This proves the convergence of (B2) because $k!$ in the denominator grows faster than the numerator. In our calculations for non-Hamiltonian systems we did not observe any indications of the divergence, either.

Although the formulae (B4, B5) look rather formal, they exactly correspond to the idea of the proposed QMC scheme. We simulate a Markov random walk in a space of all possible $K$ with a probability density $P_K \propto |\Omega_K|$ to visit each state. Two kinds of trial steps are necessary: one should try either to increase or to decrease $k$ by 1, and, respectively, to add or to remove the corresponding tetrad of "coordinates". Then the standard Metropolis acceptance criterion can be constructed using the ratio

$$\frac{||w||}{k+1} \cdot \left| \frac{D_{r_1'...r_{2k+2}'}^{r_1...r_{2k+2}}}{D_{r_1'...r_{2k}'}^{r_1...r_{2k}}} \right| \qquad (B6)$$

for the incremental steps and its inverse for the decremental ones. In general, one may also want to add/remove several tetrades simultaneously. Corresponding probabilities can be constructed in a way similar to (B6).

The most time consuming operation of the algorithm is the calculation of the ratio of determinants and Green-function matrices. It is necessary for the calculation of MC weights, as well as for the Green's functions. However, there also exist so called fast-update formula for the calculation of the ratios of determinants and Green-function matrices. The usual procedure takes $N^3$ operations, while the fast-update technique requires only $N^2$ or less operations, where $N$ is the matrix size. The only matrix which one needs to keep during QMC steps is the inverse matrix of the bare Green's functions: $M^{(2k)} = (g^{(2k)})^{-1}$. The fast-update formulas for the increment of the matrix size by 1 are as follows:

$$M_{i,j}^{(2k+1)} = M_{i,j}^{(2k)} - \lambda^{-1} \left( \sum_{i'} M_{i,i'}^{(2k)} g_{i'}^{2k+1} \right) \left( \sum_{j'} g_{2k+1}^{j'} M_{j'j}^{(2k)} \right), \qquad (B7)$$

$$\lambda = 1 + \sum_{i'j'} g_{2k+1}^{j'} M_{j'i'}^{(2k)} g_{i'}^{2k+1},$$

as can be checked by matrix multiplication of (B7) with $g^{(2k+1)}$. Here, the indices take the values $1..2k+1$, and the matrix $M_{(2k)}$ is enlarged to a $(2k+1) \times (2k+1)$ matrix with $M_{2k+1,2k+1} = 1$ and $M_{2k+1,i} = 0$, $M_{i,2k+1} = 0$. The factor $\lambda$ appears to be exactly the ratio of the determinants. An incremental step $k \to k+1$ consists of the two updates (B7). One can see that the formula indeed have an $N^2$ complexity.

A decrement of the matrix size is similar:

$$M_{i,j}^{(2k-1)} = M_{i,j}^{(2k)} - \lambda^{-1} \left( \sum_{i'} M_{i,i'}^{(2k)} g_{i'}^{2k} \right) \left( \sum_{j'} g_{2k}^{j'} M_{j'j}^{(2k)} \right), \qquad (B8)$$

$$\lambda = 1 + \sum_{i'j'} g_{2k}^{j'} M_{j'i'}^{(2k)} g_{i'}^{2k}.$$

It can be shown that for the decrement step $\lambda$ in fact equals the simple ratio $M_{2k\ 2k}^{(2k)}/g_{2k}^{2k}$.

With the matrix $M$, the Green's function can be obtained both in imaginary time and at Matsubara frequencies:

$$G^\tau_{\tau'} = g^\tau_{\tau'} - \sum_{i,j} g^\tau_{\tau_i} M_{i,j} g^{\tau_j}_{\tau'} \tag{B9}$$

$$G(\omega) = g(\omega) - g(\omega)[\frac{1}{\beta} \sum_{i,j} M_{i,j} e^{i\omega(\tau_i - \tau_j)}] g(\omega).$$

Here $g^\tau_{\tau'}$ and $g(\omega)$ are the bare Green's function in imaginary time and Matsubara spaces, respectively.

In order to reduce and in some cases avoid the sign problem in CT-QMC we introduce additional quantities $\alpha^r_{r'}$ which, in principle, can be functions of time, spin and the number of lattice states. Thus, up to an additive constant we have the new separation of our action:

$$S_0 = \int\int \left( t^{r'}_r + \int\int \alpha^{r_2}_{r'_2}(w^{r'r'_2}_{rr_2} + w^{r'_2 r'}_{r_2 r}) dr_2 dr'_2 \right) c^\dagger_{r'} c^r \, dr dr',$$

$$W = \int\int\int\int w^{r'_1 r'_2}_{r_1 r_2} (c^\dagger_{r'_1} c^{r_1} - \alpha^{r_1}_{r'_1})(c^\dagger_{r'_2} c^{r_2} - \alpha^{r_2}_{r'_2}) dr_1 dr'_1 dr_2 dr'_2.$$

In this case, the determinants $D^{r_1...r_{2k}}_{r'_1...r'_{2k}}$ of $2k \times 2k$ matrices have the following form:

$$D^{r_1 r_2 ... r_{2k}}_{r'_1 r'_2 ... r'_{2k}} = \det |g^{r_i}_{r'_j} - \alpha^{r_i}_{r'_j} \delta_{ij}| \tag{B10}$$

where $i, j = 1, ..., 2k$. A proper choice of $\alpha$ can completely remove the sign problem. For example, in the case of Hubbard model, the choice $\alpha_\uparrow = 1 - \alpha_\downarrow$ eliminates the sign problem for repulsive systems with particle-hole symmetry. Note however, that the proper choice of the $\alpha$'s depends on the particular system under consideration. For the calculation presented in this article we use the symmetrized expression for the interaction:

$$W = \frac{1}{2} \sum_{mm'\sigma\sigma'} (U_{mm'} - \delta_{\sigma\sigma'} J_{mm'})(n^\sigma_m - \alpha)(n^{\sigma'}_{m'} - 1 + \alpha). \tag{B11}$$

With this form of interaction, a small negative $\alpha$ was found to eliminate the sign problem completely. Value of $\alpha = -0.01$ was used in the calculation.

Let us now describe how the global updates can be implemented in CT-QMC calculation for the insulating phase. This trick is useful to get rid of the local trapping of the Markov random walk for the system with reasonably small orbital or spin polarization, specially for insulating systems. Another possibility to eliminate local trapping would be to use Quantum Wang-Landau procedure[56]. Although it is also implemented in CT-QMC code, we realized that global updates are practically much more efficient for our case.

To analyze the structure of perturbation series (B5), it is useful to consider an atomic limit $\mathcal{G}^{-1} = i\omega + \mu$. In this case, the last line of (B2) is composed of particle-number operators which actually do not depend on time argument, because $n_{s\sigma}$ commutes with non-interacting part of the action. Therefore the average (B10) can be calculated according to the formula $\langle (n^\uparrow_1 - \alpha)^{k_1} (n^\uparrow_1 - 1 + \alpha)^{k'_1} ... (n^\downarrow_3 - \alpha)^{k_6} (n^\downarrow_3 - 1 + \alpha)^{k'_6} \rangle_0$ with certain power indices $k_1, k'_1, ..., k_6, k'_6$. Since for Fermi operators $n = n^2$, it can be easily shown that $(n^\sigma_m - \alpha)^k (n^\sigma_m - 1 + \alpha)^{k'} = (-\alpha)^k (\alpha - 1)^{k'}(1 - n^\sigma_m) + (1 - \alpha)^k \alpha^{k'} n^\sigma_m$. Coefficients here fall down exponentially as $k$ and $k'$ grow. Therefore, qualitatively speaking, the expression is remarkably large if either $k$ or $k'$ equal zero. We conclude that the partition function is at most contributed by such terms that 6 of 12 power indices vanish. Other six indices are large; their typical value in calculation is $k \propto \beta U \gg 1$ (Ref.[54]). In an elementary CT-QMC move, value of $k$ is changed $k + 1$ ar $k - 1$. Clearly, it is hard to switch, for example, from the situation $k_1 \gg 1, k'_1 = 0$ to $k_1 = 0, k'_1 \gg 1$, and therefore program becomes trapped in a configuration with certain set of vanished indices. It can be shown that an exponential barrier separates the traps. Practically, this results in an incorrect calculation output with nonphysically large spin and/or orbital polarization.

DMFT calculation for test systems convinced us that the trapping remains important also far from atomic limit, until the system is in the insulating state. For metallic phase, the barriers disappear.

To get rid of trapping, we introduce a special kind of Monte-Carlo steps, so called global updates for orbital and spin indices. During global update, program permutes randomly orbital and spin indices of the current realization: $m \to \tilde{m}, \sigma \to \tilde{\sigma}$. After this the weight of a new configuration should be calculated. This requires calculation of the factor $\langle (n^{\tilde{\sigma}_1}_{\tilde{m}_1}(\tau_1) - \alpha)(n^{\tilde{\sigma}'_1}_{\tilde{m}'_1}(\tau_1) - 1 + \alpha) \times ... \times (n^{\tilde{\sigma}_k}_{\tilde{m}_k}(\tau_k) - \alpha)(n^{\tilde{\sigma}'_k}_{\tilde{s}'_k}(\tau_k) - 1 + \alpha) \rangle_0$. All other quantities affecting the weight of the configuration remain the same, because the interaction part is symmetrical with respect to permutations of spin and orbital indices. Moves are accepted according to a standard Metropolis criterion. Without orbital splitting, all local minima are equivalent, therefore a global update actually does not change the weight factor, and the move is always accepted. For a small orbital field, the acceptance rate of the global moves remains high enough.

During the global update, averages over the Gaussian part of the action should be calculated from scratch. Therefore, a global update requires $\propto k^3$ operations. The operation count for elementary CT-QMC move is $\propto k^2$, because of the fast-update trick. Therefore, we perform global updates much more rarely than the elementary CT-QMC move. This turns out to be sufficient to get rid of the local trapping.

---

[1] For a review, see: N. F. Mott, 1990, Metal-Insulator Transitions (Taylor and Francis, London/Philadelphia). W. Brückner, H. Oppermann, W. Reichelt, J. I. Terukow, F. A. Tschudnowski, and E. Wolf, 1983, *Vanadiumoxide Darstellung, Eigenschaften, Anwendung* (Akademie, Berlin).

[2] D.B. McWhan and T.M. Rice, Phys. Rev. Lett. **22**, 887 (1969); D.B. McWhan, T.M. Rice, and J.P. Remeika, Phys. Rev. Lett. **23**, 1384 (1969).

[3] H. Kuwamoto, J.M. Honig, and J. Appel, Phys. Rev. B **22**, 2626 (1980).

[4] G.A. Thomas et al., Phys. Rev. Lett. **73**, 1529 (1994).

[5] S. Shin, S. Suga, M. Taniguchi, M. Fujisawa, H. Kanzaki, A. Fujimori, H. Daimon, Y. Ueda, K. Kosuge, and S. Kachi, Phys. Rev. B **41**, 4993 (1990); S. Shin, Y. Tezuka, T. Kinoshita, T. Ishii, T. Kashiwakura, M. Takahashi, and Y. Suda, J. Phys. Soc. Jpn. **64**, 1230 (1995).

[6] G. A. Sawatzky and D. Post, Phys. Rev. B **20**, 1546?1555 (1979).

[7] M. Schramme, Ph.D. thesis, Universität Augsburg, 2000; E. Goering, M. Schramme, O. Müller, H. Paulin, M. Klemm, M.L. denBoer, S. Horn, Physica B **230-232**, 996 (1997).

[8] S.-K. Mo, J. D. Denlinger, H.-D. Kim, J.-H. Park, J. W. Allen, A. Sekiyama, A. Yamasaki, K. Kadono, S. Suga, Y. Saitoh, T. Muro, P. Metcalf, G. Keller, K. Held, V. Eyert, V. I. Anisimov, D. Vollhardt, Phys. Rev. Lett. **90**, 186403 (2003).

[9] S.-K. Mo et al., Phys. Rev. Lett. **93**, 076404 (2004).

[10] D.B. McWhan, J.P. Remeika, T.M. Rice, W.F. Brinkman, J.P. Maita, and A. Menth, Phys. Rev. Lett. **27**, 941 (1971).

[11] C. Castellani, C.R. Natoli and J. Ranninger, Phys. Rev. B **18**, 4945 (1978); *ibid* **18**, 5001 (1978).

[12] M. J. Rozenberg et al., Phys. Rev. Lett. **75**, 105 (1995).

[13] J.-H. Park et al., Phys. Rev. B **61**, 11506 (2000).

[14] F. Mila, R. Shiina, F.C. Zhang, A. Joshi, M. Ma, V.I. Anisimov, and T.M. Rice, Phys. Rev. Lett **85**, 1714 (2000).

[15] S. Di Matteo, N. B. Perkins, and C. R. Natoli, Phys. Rev. B **65**, 054413 (2002).

[16] L.F. Mattheiss, J. Phys.: Condens. Matter **6**, 6477 (1994).

[17] S. Ezhov, V. Anisimov, D. Khomskii and G. Sawatzky, Phys. Rev. Lett. **83**, 4136 (1999).

[18] I. S. Elfimov, T. Saha-Dasgupta, and M. A. Korotin, Phys. Rev. B **68**, 113105 (2003).

[19] V. Eyert, U. Schwingenschlögl, and U. Eckern, Europhys. Lett. **70**, 782-788 (2005).

[20] K. Held, G. Keller, V. Eyert, D. Vollhardt, and V.I. Anisimov, Phys. Rev. Lett. **86**, 5345 (2001).

[21] G. Keller, K. Held, V. Eyert, D. Vollhardt, and V.I. Anisimov, Phys. Rev. B **70**, 205116 (2004); D. Vollhardt et al., J. Phys. Soc. J. **74**, 136 (2005).

[22] U.K. Poulsen, J. Kollar, and O.K. Andersen, J. Phys. F **6**, L241 (1976).

[23] G. Sawatzky, recorded talk at the Kavli Institute for Theoretical Physics, UCSB, Nov 2002.

[24] V.I. Anisimov, D.E. Kondakov, A.V. Kozhevnikov, I.A. Nekrasov, Z.V. Pchelkina, J.W. Allen, S.-K. Mo, H.-D. Kim, P. Metcalf, S. Suga, A. Sekiyama, G. Keller, I. Leonov, X. Ren, and D. Vollhardt, Phys. Rev. B **71**, 125119 (2005). In order to avoid the fixed hybridization approximation one needs to carry out the full **k**-sum in the self-consistency condition (Eq. 8). This can either be done by using the full set of nearly orthonormal LMTOs corresponding to the valence orbitals, as has been done in transition metals[25,44] but is technically involved for transition metal oxides (due the strong hybridization), or the set must be downfolded to $t_{2g}$ Wannier functions. The above-mentioned paper presents a way of doing this. The first consistent use of the Wannier representation in LDA+DMFT calculations was by Pavarini et al.[29].

[25] M.I. Katsnelson, A.I. Lichtenstein, J. Phys. Cond. Mat. 11 (4): 1037 (1999); A.I. Lichtenstein, M.I. Katsnelson, G. Kotliar, Phys. Rev. Lett. 87 (6): 067205 (2001).

[26] M. S. Laad, L. Craco, and E. Müller-Hartmann, Phys. Rev. B **73**, 045109 (2006); M. S. Laad, L. Craco, and E. Müller-Hartmann, Phys. Rev. Lett. **91**, 156402 (2003).

[27] T. Saha-Dasgupta, J. Nuss, O. K. Andersen, A. I. Lichtenstein, A. I. Poteryaev, and A. Georges, preceding paper.

[28] O.K. Andersen and T. Saha-Dasgupta, Phys. Rev. B **62**, R16219 (2000) and references therein.

[29] E. Pavarini, S. Biermann, A.I. Poteryaev, A.I. Lichtenstein, A. Georges, and O. K. Andersen, Phys. Rev. Lett. **92**, 176403 (2004); E. Pavarini, A. Yamasaki, J. Nuss, and O.K. Andersen, New J. Phys. **7**, 188 (2005).

[30] N. Mazari and D. Vanderbilt, Phys. Rev **B 56** 12847 (1997).

[31] F. Lechermann, A. Georges, A. Poteryaev, S. Biermann, M. Posternak, A. Yamasaki, and O.K. Andersen, Phys. Rev. B **74**, 125120 (2006).

[32] R. Frésard and G. Kotliar, Phys. Rev. B **56**, 12909 (1997).

[33] S. Savrasov, G. Kotliar and E. Abrahams, Nature **410** 293 (2001); L. Chioncel, L. Vitos, I. Abrikosov, J. Kollár, M.I. Katsnelson, and A.I. Lichtenstein, Phys. Rev. B **67** 235106 (2003); J. Minár, L. Chioncel, A. Perlov, H. Ebert, M.I. Katsnelson, and A.I. Lichtenstein, Phys. Rev. B **72** 045125 (2005).

[34] I. Solovyev, N. Hamada, and K. Terakura, Phys. Rev. B **53**, 7158 (1996).

[35] A. Georges and G. Kotliar, Phys. Rev. B **45**, 6479 (1992); H. Kajueter and G. Kotliar, Phys. Rev. Lett. **77**, 131-134 (1996).

[36] T. Pruschke and N. Grewe, Z. Phys. B **74**, 439 (1989); M. Jarrell and T. Pruschke, Z. Phys. B **90**, 187 (1993); M. Jarrell and T. Pruschke, Phys. Rev. B **49**, 1458 (1993).

[37] M. Jarrell and J.E. Gubernatis, Physics Reports **269**, 133 (1996).
[38] J.M. Tomczak and S. Biermann, to be published.
[39] Spectral functions along other directions available upon request to Jan.Tomczak@cpht.polytechnique.fr.
[40] M. Jarrell, J. K. Freericks, and Th. Pruschke, Phys. Rev. B **51**, 11704 (1995).
[41] N. Blümer, PhD thesis, Universität Augsburg (2002).
[42] I.A. Nekrasov *et al.*, Phys. Rev. B **73**, 155112 (2006).
[43] A.I. Lichtenstein and M.I. Katsnelson, Phys. Rev. B **57**, 6884 (1998).
[44] S. Biermann, A. Dallmeyer, C. Carbone, W. Eberhardt, C. Pampuch, O. Rader, M.I. Katsnelson and A.I. Lichtenstein, JETP Letters **80**, 612 (2004) and cond-mat/0112430.
[45] A. Georges, G. Kotliar, W. Krauth and M. Rozenberg, Rev. Mod. Phys. **68**, 13 (1996)
[46] I. Turek, V. Drchal, J. Kudrnovsky, M. Sob, and P. Weinberger, Electronic Structure of Disordered Alloys, Surfaces and Interfaces, (Kluwer Academic Publishers, Boston, 1997)
[47] A. Yamasaki *et al.*, Phys. Rev. Lett. **96**, 166401 (2006).
[48] T. Saha-Dasgupta, A. Lichtenstein and R. Valenti, Phys. Rev. B **71**, 153108 (2005).
[49] J.E. Hirsch and R.M. Fye, Phys. Rev. Lett. **56**, 2521 (1986).
[50] J.E. Hirsch, Phys. Rev. B **28**, 4059 (1983).
[51] A.I. Poteryaev, A.I. Lichtenstein and G. Kotliar Phys. Rev. Lett. **93**, 086401 (2004).
[52] S. Biermann, A.I. Poteryaev, A.I. Lichtenstein, A. Georges, Phys. Rev. Lett. **92**, 026404 (2005).
[53] A. Sekiyama *et al.* Phys. Rev. Lett. **93**, 156402 (2004).
[54] A.N. Rubtsov, cond-mat/0302228; A.N. Rubtsov and A. I. Lichtenstein, JETP Lett. **80**, 61 (2004); A.N. Rubtsov, V.V. Savkin, and A.I. Lichtenstein, Phys. Rev. B **72**, 035122 (2005).
[55] We perform the fit below the coherence temperature where we can use the zero temperature form of the self-energy.
[56] M. Troyer, S. Wessel, F. Alet, Phys. Rev. Lett. **90**, 120201 (2003).
[57] It is easy to see that the construction using the unhybridized band structure is better than using the full LDA band structure as long as the crystal field splitting of the former (that is the difference of the local LDA levels) is smaller than the hybridization. In that case, the construction using the unhybridized bands coincides with the real quasi-particle band structure up to second order in the hybridization whereas the usage of the LDA induces a first order error.
[58] We note that, strictly speaking, it is *not enough* that the crystal field splitting exceeds the value of $\max \epsilon_{e_g^\pi} - \min \epsilon_{a_{1g}}$, but one has to have separately $\min \epsilon_{a_{1g}} + \Re\Sigma_{a_{1g}}(0) > 0$ and $\max \epsilon_{e_g^\pi} + \Re\Sigma_{e_g^\pi}(0) < 0$ that is, one has to push the top of the $e_g^\pi$ band below and the bottom of the $a_{1g}$ band above the Fermi level. This is only roughly the same condition, since due to the largely incoherent character of the $e_g^\pi$ bands, Luttinger's theorem does not hold.